\documentclass[11pt]{article}

\usepackage[margin=1in]{geometry}
\usepackage{amsmath,amssymb,amsthm}
\usepackage{graphicx}
\usepackage{booktabs}
\usepackage{dcolumn}    
\usepackage{multirow}
\usepackage[round]{natbib}     
\usepackage[bookmarksopen,bookmarksnumbered,colorlinks=true,linkcolor=blue,citecolor=blue,urlcolor=blue]{hyperref}

\newcolumntype{d}[1]{D{.}{.}{#1}}

\theoremstyle{plain}

\title{Structure-Preserving Visualization of Complex Systems through Discrete Approximation: An Application to Argo Data}

\author{Shang-Ying Shiu$^{a}$, \quad Fushing Hsieh$^{b}$, \quad Ting-Li Chen$^{c,}$\thanks{Corresponding author. Email: \href{mailto:tlchen@stat.sinica.edu.tw}{tlchen@stat.sinica.edu.tw}}\\[1ex]
{\small $^{a}$Department of Statistics, National Taipei University, Taiwan}\\
{\small $^{b}$Department of Statistics, University of California, Davis, California, USA}\\
{\small $^{c}$Institute of Statistical Science, Academia Sinica, Taiwan}}

\date{}

\begin{document}

\maketitle

\begin{abstract}
This paper presents a framework for constructing structure-preserving representations of complex systems through discrete approximation, and demonstrates its use in studying the vertical temperature and salinity structures in the mesopelagic zone across the global ocean using the ARGO dataset. Clustering serves as a means of organizing complexity into a finite set of structures that approximate the overall oceanic conditions, and a color encoding design then integrates these structures into a coherent map, with the three color components derived from interpretable geometric features of a profile: its initial level, its magnitude of variation, and its shape. Instead of focusing on specific depth levels or computing zonal averages within selected regions, our approach preserves the full vertical structure of individual profiles and incorporates each profile in the global ocean, capturing both fine-scale profile detail and large-scale spatial variability. By clustering over one million profiles collected over a decade, we identify and characterize representative profile shapes, which form the basis for a visualization strategy that provides an integrated, comprehensive, and interpretable presentation of the large-scale spatial distributions of these oceanic vertical patterns.
\end{abstract}


\section{Introduction}
\label{sec:intro}
Many large-scale systems, such as natural environments or socio-economic networks, exhibit spatial variation that is expressed through high-dimensional profiles. These systems are often complex and potentially continuous in nature, posing challenges to study and interpret them. This paper presents a framework for studying such vast and complex systems. The framework was developed through the analysis of ARGO data \citep{argo2000argo} to understand the diverse temperature and salinity profile characteristics across different regions of the global ocean. The major difficulty in this analysis is the high complexity of ocean structures, which do not conform to a small number of patterns and instead exhibit gradual transitions across regions. This property makes it difficult to provide a comprehensive and interpretable representation of both global patterns and localized variations, motivating the development of this framework to characterize and visualize large-scale spatial complexities.

The framework begins with the use of clustering, a fundamental tool in unsupervised learning that groups similar observations based on intrinsic patterns. Clustering enables researchers to identify meaningful subgroups without requiring labels or strong parametric assumptions. In conventional settings, data are assumed to arise from a finite set of discrete groups, and the goal is to discover and interpret those groups.

The role of clustering, however, need not be confined to discovering discrete groups. There is a broader and less conventional perspective that views clustering as a tool for dimension reduction, which is especially relevant when dealing with complex objects such as curves and profiles. While conventional dimension reduction methods, such as principal component analysis (PCA; \citealt{pearson1901liii}), project high-dimensional data into a continuous low-dimensional space, clustering instead simplifies this data into a categorical one-dimensional variable, where each observation is represented by a cluster label. This functional role of clustering, as a means of representing high-dimensional data through discrete prototypes, offers an alternative approach beyond conventional continuous dimension reduction techniques. For readers accustomed to thinking of clustering purely as grouping, this reinterpretation reveals a hidden but powerful function of the method: it enables characterization by structure, not just separation by similarity.

This perspective can be illustrated through our analysis of Argo data, which contains temperature and salinity measurements at various depths throughout the global ocean. Common practices for analyzing oceanographic vertical patterns include summarizing over fixed depth layers, which often fail to capture important vertical structures, or averaging within regions, which makes it difficult to observe spatial variations. To address these limitations, we adopt an alternative approach that retains the full shape of the temperature and salinity profiles and applies clustering globally to identify representative vertical patterns. This approach reduces the high-dimensional temperature and salinity profiles to a one-dimensional cluster label while preserving the essential structural characteristics, as captured by the representative profile of each cluster.

Pushing this idea further, clustering can go beyond characterizing structures to function as a means of approximating continuity in situations where no discrete partitions can be meaningfully interpreted as clusters. This can be seen in the analysis of the Argo data, where temperature and salinity profiles often exhibit smooth transitions across regions, rather than clearly defined boundaries. In such settings, clustering is not used to detect conventional cluster structures, but to provide a discretized and interpretable map of variation, with each cluster serving as a local anchor summarizing a region. In this context, clustering is not the destination, but the path that helps organize complexity when traditional boundaries do not exist. It offers a structured approximation that captures rich, continuous variation through a finite set of representative characteristics.

Given this approximation, however, presenting the global spatial patterns requires more than just identifying the representative characteristics by clustering. It also calls for a structured way to convey the information each cluster holds and to integrate the information into a coherent global representation, rather than treating the clusters as isolated elements. To enable such a representation, we design a cluster-based color encoding using the OKLCH color model. The encoding reflects the relationships among cluster-representative characteristics, thereby allowing their similarities and differences to be intuitively perceived and highlighting the smooth transitions between these characteristics. This color encoding, together with clustering, constitutes the framework applied to Argo data, revealing both the fine-scale vertical structure present in individual profiles and the large-scale spatial variation that becomes apparent only through global clustering.

In the remainder of this article, we detail the analysis process and develop a representation of global oceanographic patterns. The structure of the paper is as follows. Section 2 introduces the dataset and preprocessing steps. Section 3 describes the clustering method and presents the clustering results. Section 4 develops a color encoding scheme for visualizing these results, which Section 5 uses to display global oceanographic patterns. Section 6 concludes the study and discusses possible directions for future analysis.

\section{Data Introduction and Preparation}
\subsection{The ARGO Observing System and Selected Data Scope} 
Traditional oceanographic data collection is often spatially and temporally biased, as it is typically restricted to accessible regions, rarely extends into the deep ocean, and is often unavailable during adverse weather or seasonal conditions. The ARGO program \citep{wong2020argo}, part of the Global Ocean Observing System, is an international program that successfully addresses these limitations with two key innovations: (1) deploying a global array of autonomous profiling floats, which continuously perform measurement cycles to collect temperature and salinity data at various depths; and (2) allowing them to drift at a depth of 1000 meters, following subsurface currents. Since its launch in 2000, the Argo array has grown significantly and now maintains more than 4,000 active floats at any given time. These floats provide extensive coverage across space, time, and depth. They are globally distributed, primarily between approximately 60°N and 60°S, operate year-round on regular cycles, and profile the ocean vertically down to 2,000 meters. These capabilities enable a consistent and widespread approach to marine data acquisition. 

The data collected by Argo floats were made freely available by the International Argo Program and the national programs that contribute to it (https://argo.ucsd.edu, https://www.ocean-ops.org). To date, Argo data have supported a wide range of scientific studies spanning multiple areas of ocean science \citep{roemmich20092004,durack2010fifty,cheng2015global,johnson2016improving}. Comprehensive reviews have documented ARGO’s transformative contributions over its first two decades \citep{riser2016fifteen,johnson2022argo}. These contributions establish ARGO as an essential resource for modern oceanographic research.

We used Argo data to examine the structure and variability of vertical profiles of temperature and salinity across the global ocean. Our analysis focuses on the mesopelagic zone (200–1000 m), which lies beneath the highly dynamic surface layer and above the relatively stable deep ocean. This layer has been relatively understudied, likely because it lacks the immediate connection to atmospheric processes and climate phenomena that characterize the surface ocean. However, it contains key features such as thermocline and halocline, which are zones of rapid change in temperature and salinity and exhibit the steepest vertical gradients with depth. These sharp transitions give rise to distinct geometric features in the profile shapes, making the vertical structure in this layer particularly interesting to characterize. Moreover, they contribute to strong density stratification, which restricts vertical mixing and mediates the exchange of heat, salt, and other properties between the surface and the deep ocean. These vertical structures are shaped by global and regional circulation patterns and thus play a central role in large-scale ocean dynamics, making this layer important to understand.

\subsection{Data Preparation}
We used delayed-mode Argo data with a quality control (QC) flag of 1, which indicates the highest level of data reliability. We extracted data from the period 2014–2023, as it represented the most recent ten years of globally available data with full yearly coverage at the time our analysis was initiated in the fall of 2024. Applying these criteria yielded a total of 9,154 floats, which collectively performed 1,260,354 cycles.

Each cycle takes measurements at different pressure levels (in decibar). To clarify, we refer to these vertical positions as depth (in meters) throughout this paper. Ensuring consistency in the observation points, we linearly interpolated all vertical profiles to a common set of 24 depth levels within the mesopelagic zone of interest, specifically at the following depths: 200, 220, 240, 260, 280, 300, 320, 340, 360, 380, 400, 420, 440, 470, 500, 550, 600, 650, 700, 750, 800, 850, 900, and 950 meters below the sea surface.

We filtered out cycles with insufficient vertical coverage to ensure consistent interpolation quality across profiles. The retained cycles were required to meet two criteria for both temperature and salinity: (1) they include measurements covering the full depth range from 150 to 1100 meters, and (2) they contain at least 20 measurements within the mesopelagic zone. After filtering, 1,059,130 cycles remained. For each cycle, we constructed interpolated temperature and salinity profiles at the 24 fixed depth levels, resulting in two datasets for subsequent clustering analysis, each consisting of 1,059,130 24-dimensional profiles.

\subsection{Profile Diversity}
Thermoclines and haloclines in the mesopelagic zone vary considerably in depth and strength across oceanic regions. As a result, the over one million interpolated profiles in each of our two datasets display a wide range of gradients and curvatures, reflecting the complex and heterogeneous nature of the intermediate ocean. In particular, unlike the temperature that almost always decreases with depth, salinity can increase with depth in certain regions before eventually approaching a value near 35 PSU in deeper waters. This leads to even greater structural complexity in salinity profiles.

\section{Clustering-Based Characterization of Global Ocean Profiles}
We perform clustering separately on 1,059,130 temperature profiles and 1,059,130 salinity profiles. This approach helps reduce the complexity of these large and high-dimensional datasets by identifying representative patterns across different regions, and it provides a simplified yet informative summary of the global distribution of temperature and salinity structures.

\subsection{Randomized SUP Clustering}
Clustering global ocean profiles presents two major challenges. First, the dataset is highly complex and diverse, reflecting a wide range of oceanographic conditions and often containing substantial noise. Second, the dataset comprises over a million profiles, posing significant computational demands. While standard clustering algorithms are limited either in their ability to handle complex and noisy data, or in their efficiency when applied to large-scale datasets, the randomized self-updating process (rSUP; \citealt{shiu2024randomized}) offers a robust and scalable solution for clustering data of this nature.

The randomized self-updating process (rSUP) is an extension of the self-updating process (SUP; \citealt{chen2007new}), which is mathematically similar to the blurring mean-shift algorithm \citep{cheng1995mean} but conceptually different. In contrast to mean-shift, where clustering is a byproduct of density mode-seeking, SUP performs clustering by iteratively updating each sample’s location in feature space according to the aggregated influence of other samples. Explicitly, 

\begin{equation}
\mathbf{x}_i^{(t+1)} = \sum_{j=1}^{N}  \mathbf{w}_{ij}^{(t)} \mathbf{x}_j^{(t)}, \nonumber
\end{equation}

\noindent where $N$ is the total number of samples in the dataset, $\mathbf{x}_i^{(t+1)}$ is the updated feature vector of sample $i$ at the $(t+1)$-th iteration, and 
\begin{equation}
\mathbf{w}_{ij}^{(t)}=\frac{f_t\left(\mathbf{x}_i^{(t)}, \mathbf{x}_j^{(t)}\right)}{\sum_{k=1}^{N} f_t\left(\mathbf{x}_i^{(t)}, \mathbf{x}_k^{(t)}\right)} \nonumber
\end{equation}

\noindent denotes the weighted influence between $\mathbf{x}_i^{(t)}$ and $\mathbf{x}_j^{(t)}$, with $f_t$ representing the influence function. This interaction resembles a gravitational force-based dynamic that gradually moves every object in the system toward an equilibrium that exhibits the intrinsic cluster structure of the data. The convergence and consistency of such processes have been studied and theoretically established \citep{chen2015convergence,chen2025mean}.

Although the self-updating process (SUP) has demonstrated strong performance on complex datasets \citep{shiu2016strengths}, its computational complexity scales quadratically with the sample size, making it inefficient for large datasets. The randomized self-updating process (rSUP) improves efficiency by introducing random sampling in each iteration. Specifically, rSUP updates each sample using a random subset instead of the full dataset, 

\begin{equation}
\mathbf{x}_i^{(t+1)} = \sum_{j \in G_t(i)}  \mathbf{w}_{ij}^{(t)} \mathbf{x}_j^{(t)}, \nonumber
\end{equation}
where $ G_t(i)$ denotes the random subset that contains sample $i$ at the $t$-th iteration. This use of random sampling significantly reduces the computational cost. In addition, the consistency property has been theoretically proven, meaning that rSUP retains the same clustering results as SUP when the size of the random subset is sufficiently large. The supc R package, which supports both SUP and rSUP, is available for download \citep{supc}.

\subsection{Clustering Procedure and Results}
Temperature and salinity profiles are considered to be functional data, because each profile can be viewed as a discretized realization of an underlying smooth function. Common approaches for clustering functional data include projecting profiles onto basis functions, such as B-splines or functional principal components, and incorporating derivatives that capture higher-order structural features. We do not take these approaches here, as nearly all profiles in our dataset exhibit a monotonic trend with depth. Instead, we consider a simpler strategy by directly measuring overall vertical differences using Euclidean distance as a measure of profile similarity. This metric is sufficient to distinguish between major structural patterns, which leads to strong clustering performance. Moreover, by avoiding derivatives, which are often unstable and require interpolation, we reduce noise that could adversely affect clustering results. 

We cluster the temperature and salinity profiles separately using the randomized self-updating process (rSUP), employing an exponentially decaying Euclidean distance to determine the strength of influence $f_t$ between $\mathbf{x}_i^{(t)}$ and $\mathbf{x}_j^{(t)}$,

\begin{equation}
f_t\left( \mathbf{x}_i^{(t)}, \mathbf{x}_j^{(t)} \right) =
\begin{cases}
\exp\left( -\dfrac{ \|\mathbf{x}_i^{(t)}-\mathbf{x}_j^{(t)} \|_2}{ T(t) } \right), & \text{if }  \|\mathbf{x}_i^{(t)}-\mathbf{x}_j^{(t)} \|_2 \leq r \\
0, & \text{if } \|\mathbf{x}_i^{(t)}-\mathbf{x}_j^{(t)} \|_2 > r, \nonumber
\end{cases}
\end{equation}

\noindent where $T(t)$ is the temperature parameter that affects the update speed, and $r$ is the influence range, defining the radius of the neighborhood beyond which samples no longer influence each other. For $T(t)$, we simply use the default setting in the supc package. For the influence range parameter $r$, our selection is based on the purpose of clustering: The resulting clusters are intended to approximate global oceanographic patterns. Therefore, the ideal value of r should produce a sufficiently fine-grained partitioning to capture subtle differences among profiles, while still ensuring that the resulting cluster representatives remain distinguishable from each other.

Based on empirical observations, we selected the influence range parameter r to be the fourth percentile of the pairwise profile distances for temperature profile clustering, resulting in 721 clusters. Among these, 497 were singleton clusters and many more with very few profiles, indicating that some profiles were identified as outliers or noise by the updating process and were therefore excluded from the formation of major clusters. We selected the top 30 clusters to represent the global vertical temperature patterns, covering 98.68\% of the temperature profiles in the dataset. On the other hand, salinity profiles are generally more diverse than temperature profiles. Consistent with this, we selected a smaller value of r, specifically the third percentile of the pairwise profile distances, for the clustering of salinity profiles. This resulted in 3,190 clusters, of which 2,462 were singleton clusters and many more with very few profiles. We selected the top 50 clusters to summarize the global vertical salinity patterns, which cover 97.26\% of the salinity profiles in the dataset.

\subsection{Cluster-Representative Profile Shapes}

\begin{figure}
\begin{center}

  \begin{minipage}{\textwidth}
    \centering
    \includegraphics[width=6.4in]{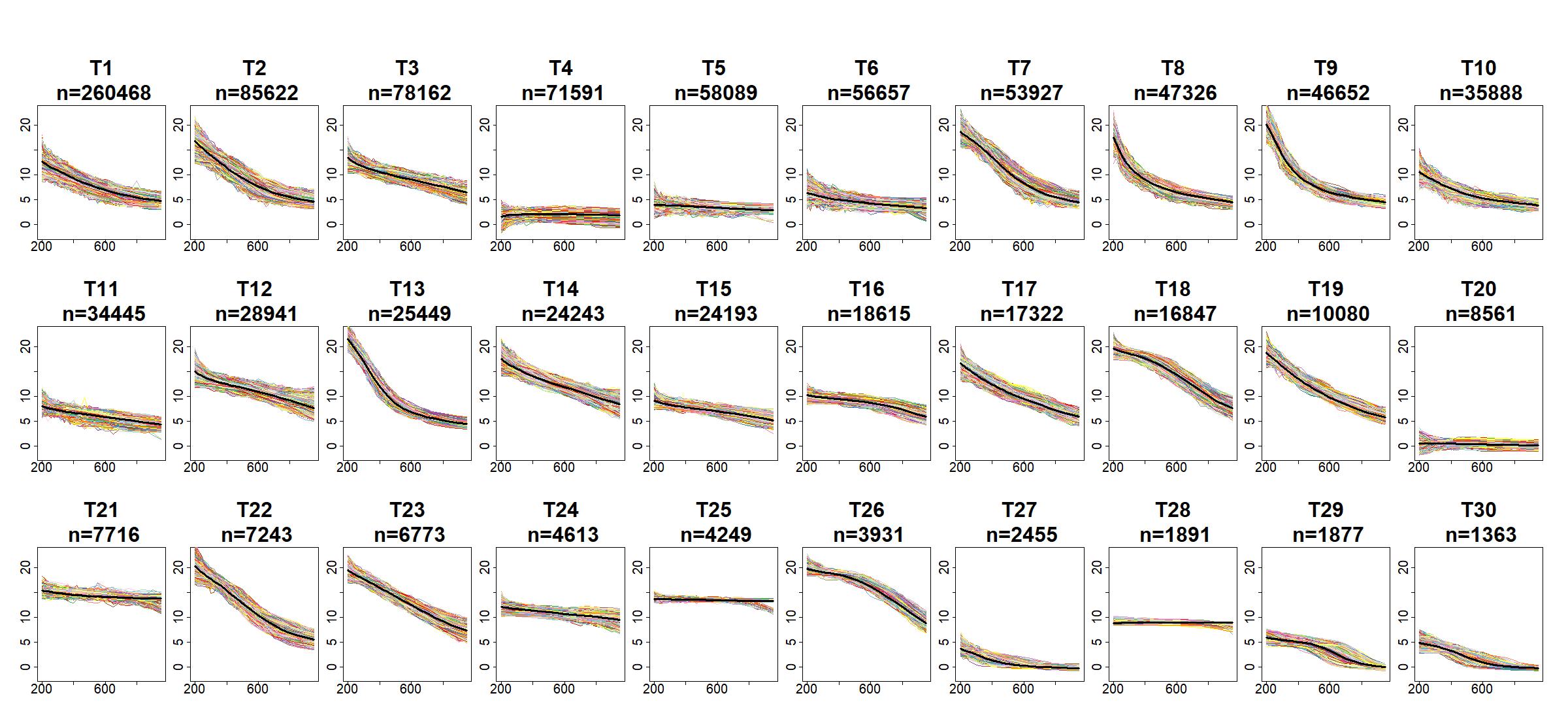}

    \vspace{-0.3cm}

    (a) Top 30 Clusters of Temperature Profiles   
  \end{minipage}

  \vspace{0cm}
  
  \begin{minipage}{\textwidth}
    \centering
    \includegraphics[width=6.4in]{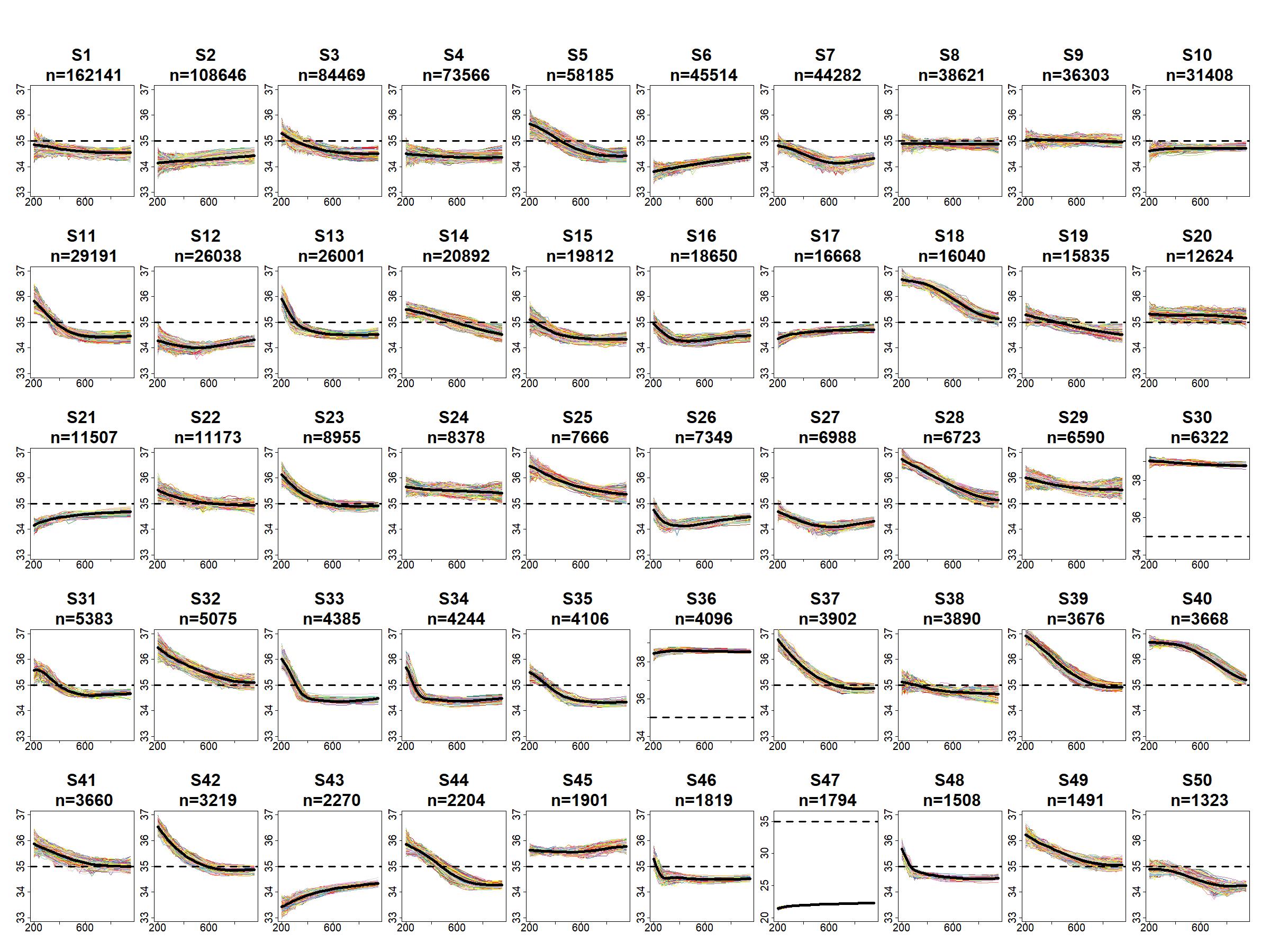}
    
    \vspace{-0.3cm}

    (b) Top 50 Clusters of Salinity Profiles   
  \end{minipage}

  \vspace{0.1cm}
  
\caption{Cluster-representative profiles. The x-axis represents depth and the y-axis denotes temperature or salinity. }
\label{fig:cluster_profiles}
\end{center}
\end{figure}

Figure~\ref{fig:cluster_profiles}(a) and~\ref{fig:cluster_profiles}(b) show the top 30 temperature clusters and the top 50 salinity clusters, respectively. The shaded regions form compact bands around the representative curves, indicating strong within-cluster homogeneity and providing visual evidence of the strong clustering performance of the rSUP algorithm.

Upon examining these cluster-representative profiles, we notice that some share very similar shapes and can be considered part of the same type, while some exhibit very distinct geometric structures. The 30 temperature profiles can be broadly classified into four shape categories: approximately linear (horizontal or with a negative slope), smooth decreasing convex, and smooth decreasing concave. In addition, the 50 salinity profiles display a greater diversity and exhibit four more shape categories, including increasing patterns such as linear with positive slope and smooth increasing convex, and more complex forms such as bending convex-decreasing and cup-shaped curves (initially decreasing, then slightly increasing). These representative shapes show how temperature and salinity vary with depth, providing a discrete set of pattern characteristics that summarize the oceanographic vertical structure across the global ocean.

\subsection{Cluster Geographic Locations and Profile Transitions}
To understand where different representative profile shapes occur, what oceanographic characteristics are associated with each region, and how these profiles transition across geographic space, we begin by examining cluster geographic locations.

\begin{figure}
\begin{center}

  \begin{minipage}{\textwidth}
    \centering
    \includegraphics[width=6in]{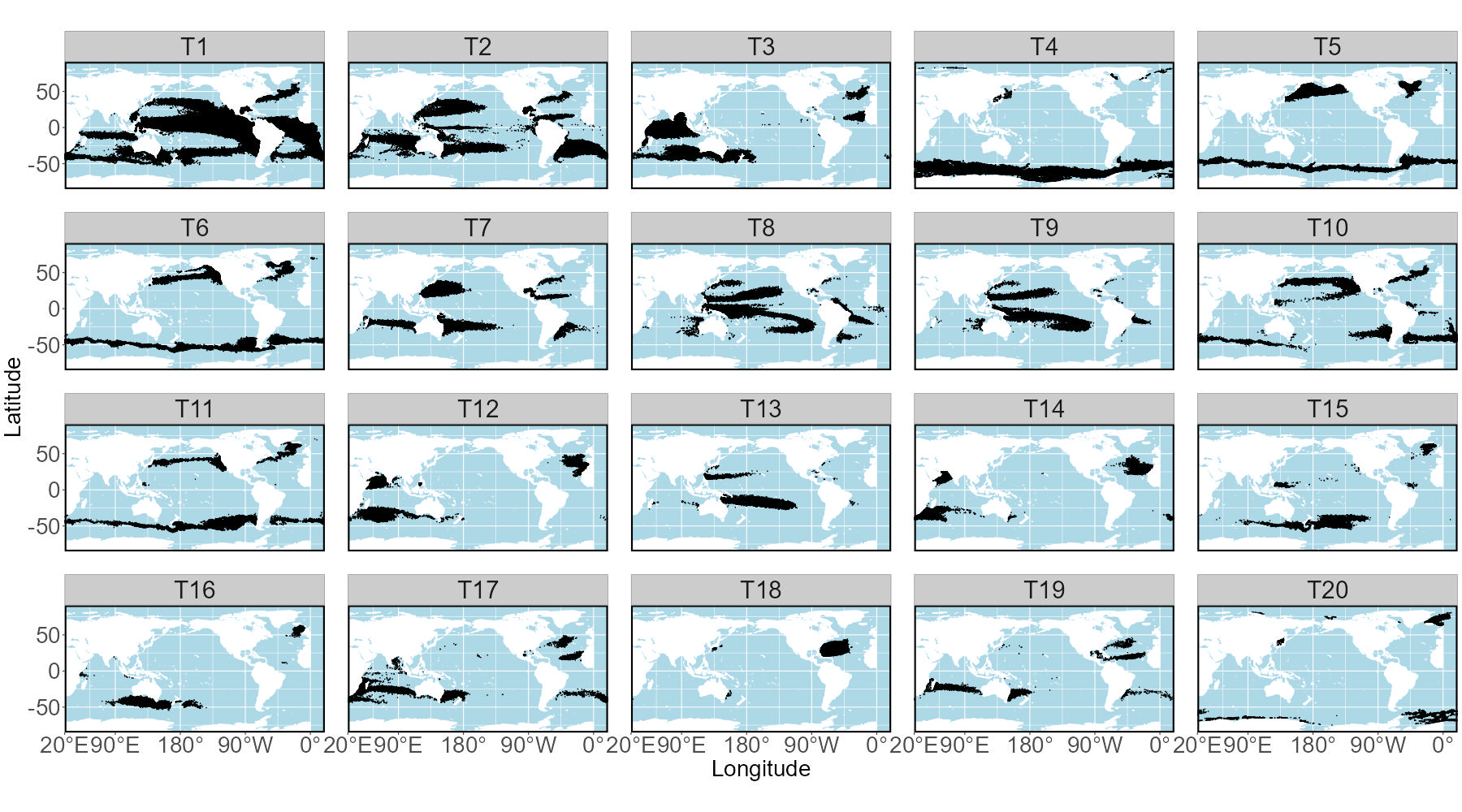}
  \end{minipage}
  
  \vspace{-0.4cm}
  \begin{minipage}{\textwidth}
    \centering
    \includegraphics[width=6in]{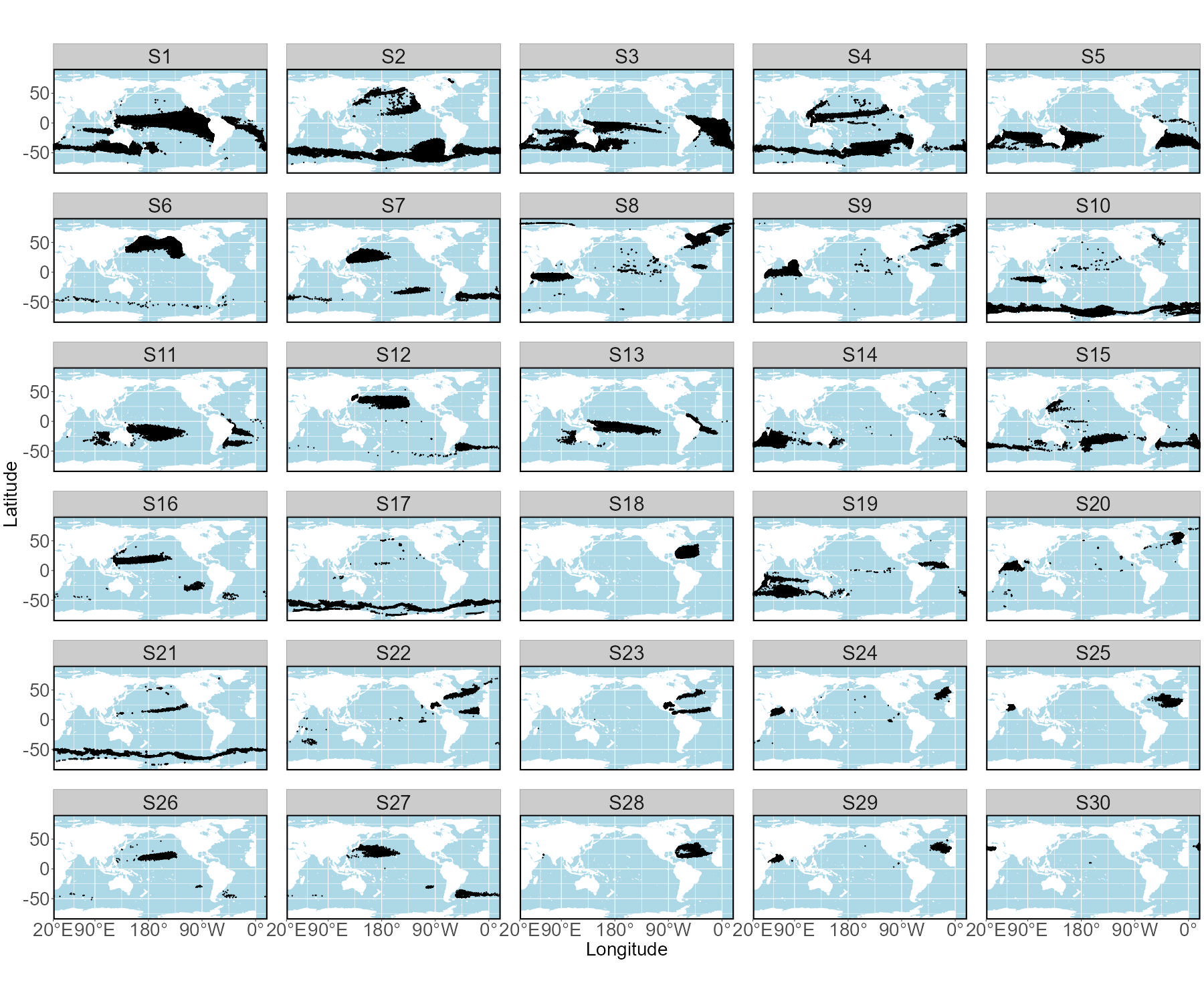}
  \end{minipage}

  \vspace{-1cm}
\end{center}
\caption{Geographic distribution of Argo cycles corresponding to the top 20 temperature clusters and top 30 salinity clusters.}
\label{fig:profile_loc}
\end{figure}

 Figure~\ref{fig:profile_loc}(a) and~\ref{fig:profile_loc}(b) illustrate the locations of cycles belonging to the top twenty temperature clusters and the top thirty salinity clusters, which account for 94\% and 91\% of the total cycles, respectively. Two notable spatial patterns can be identified. First, despite being defined solely by profile shapes, the clusters exhibit spatial grouping. Second, regions associated with different clusters are not strictly separated and often overlap, meaning that cycles observed at nearby locations may belong to different clusters.  

\begin{figure}
\begin{center}

  \begin{minipage}{\textwidth}
    \centering

    \includegraphics[width=6.4in]{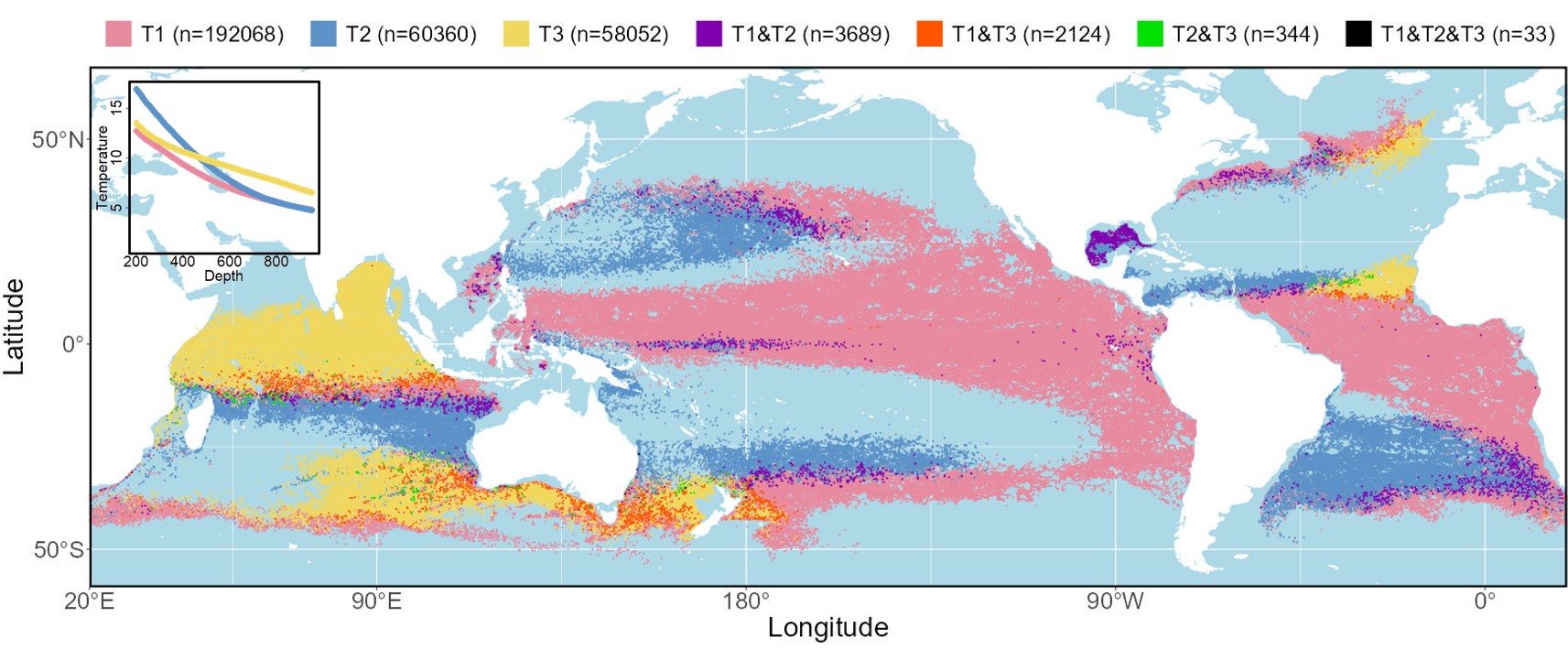}  
    
    \vspace{-0.1cm}
    
    \textbf{(a)} Geographic Distribution of T1–T3 and Their Overlaps   
  \end{minipage}

  \vspace{1cm} 
  
  \begin{minipage}{\textwidth}
    \centering
    \includegraphics[width=6.4in]{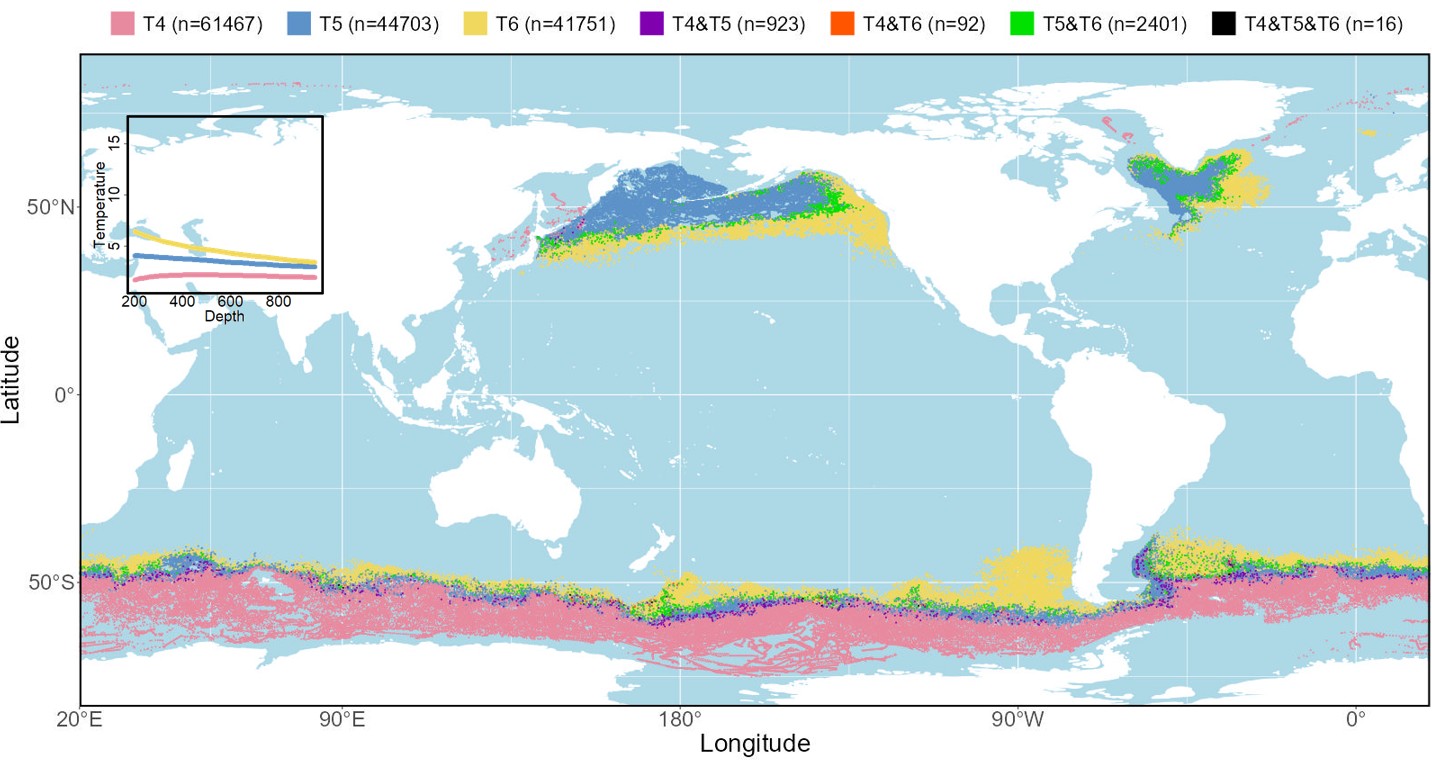}

    \vspace{-0.1cm}
    
    \textbf{(b)} Geographic Distribution of T4–T6 and Their Overlaps
  \end{minipage}

\vspace{0.4cm}
\caption{Geographic distribution of selected temperature clusters and their overlapping regions. Grids containing cycles belonging exclusively to a single cluster are colored red, blue, or yellow. Grids showing the coexistence of cycles from multiple clusters (i.e., the overlapping regions) are colored purple, orange, green, and black. The value n in the legend represent the number of grid cells associated with each cluster or cluster overlap. The inset panel in the upper-left corner displays the representative temperature profiles for the clusters.}
\label{fig:worldmap_temp_c1-c6}
\end{center}
\end{figure}

Figure~\ref{fig:worldmap_temp_c1-c6}(a) and \ref{fig:worldmap_temp_c1-c6}(b) further illustrate these two spatial patterns using grid-based maps. Both figures provide a resolution of 0.1° × 0.1°. Each grid cell is colored according to the clusters assigned to the cycles observed at that location, regardless of sampling time. The colors indicate whether the cycles in a grid belong exclusively to a single cluster or to multiple clusters. In the latter case, the grid is identified as an overlap region. Note that an overlap does not imply that different clusters were observed simultaneously within the same grid, and this definition allows us to characterize the full range of profile types ever observed in each grid.

Figure~\ref{fig:worldmap_temp_c1-c6}(a) displays the grid-based geographic distribution of temperature clusters T1, T2 and T3, showing that the three clusters occupy broad but distinct latitude bands. Clusters T1 and T3 dominate in different tropical regions, with some extension into subtropical and even temperate zones. In contrast, T2 is primarily distributed in subtropical regions, with only a small presence in the tropics. Their representative profile shapes reflect the vertical temperature structure in their respective locations. The profiles of T1 and T3 are characterized by relatively weak temperature gradients between 200 and 1000 meters, while the profile of T2 shows a higher temperature at 200 meters and exhibits much stronger vertical gradients.

Figure \ref{fig:worldmap_temp_c1-c6}(a) also highlights regions of overlap, where cycles within the same grid belong to different clusters. For example, the purple areas indicate the coexistence of both gradual (T1) and sharp (T2) changes in vertical temperature structure. These areas may be interpreted as transitional zones where the water column shifts from gradual to more strongly stratified thermal profiles. They may also reflect greater spatial heterogeneity in the vertical profiles. To illustrate, two regions in the Atlantic Ocean stand out for their diversity of colors, and both are located close to confluence zones where warm and cold currents interact. One is the Gulf Stream extension, where the warm Gulf Stream meets the cold Labrador Current. The other is located in the equatorial Atlantic, where the warm North Equatorial Current joins the cool Canary Current. These confluence zones contribute to high variability in vertical profiles, resulting in the observed spatial overlap among clusters.

Figure~\ref{fig:worldmap_temp_c1-c6}(b) provides another example that illustrates profile locations and transitions. Temperature clusters T4, T5, and T6 are primarily located in high-latitude regions, including the subpolar North Pacific and Atlantic, as well as along the Antarctic continental margins. Their latitudinal distributions differ only slightly, and their representative profiles, which correspond well to their geographic locations, also differ only slightly. T4 occurs in the highest southern latitude,  and its representative profile exhibits uniformly low temperatures with a slight increase with depth. This pattern suggests that surface waters in very high-latitude regions are likely colder than the underlying layers. T6 is found in regions furthest from the poles among the three clusters, and its profile exhibits slightly higher temperatures with a gradual decrease with depth. This decrease is marginally more pronounced than that in T5, which lies between T4 and T6 both in terms of geographic distribution and profile shape.

The purple and green bands located in the Southern Ocean near Antarctica indicate overlap between clusters. Together with the three individual clusters, they form a sequence, represented in order by red, purple, blue, green, and yellow, that extends from the coldest regions adjacent to the Antarctic continent toward relatively warmer waters farther north. This sequence corresponds to a gradual transition from T4 to T6 in both temperature magnitude and profile shape: (1) The overall temperature remains low but gradually increases from south to north; (2) The profile is nearly uniform, gradually shifting from a slight increase to a slight decrease with depth. These transitions may reflect a smooth continuum in both temperature magnitude and profile shape across these high-latitude waters in the southern hemisphere.

\section{Color Encoding of Cluster Profiles}
\label{sec:color}

The observed progression from T4 to T6 in the Southern Ocean reflects gradual changes in the underlying physical processes rather than abrupt boundaries, suggesting that the finite set of cluster profiles only serves to summarize regionally coherent structures. To allow these finite cluster profiles to approximate the inherently gradual transitions present worldwide, we develop a color encoding strategy that represents not only the geometric characteristics of the cluster profiles but also the transitions among profiles with similar structures. Such an encoding facilitates the global visualization presented in the next section.

One effective approach is to assign similar colors to profiles with similar geometric characteristics, and distinct colors to profiles with distinct shapes. Since most commonly used color models represent colors by combining three components, we use the geometric features of each profile as inputs to these components. Each profile therefore corresponds to a unique color reflecting its structural characteristics. In the following, we propose such a color-encoding strategy that produces a coherent global view of oceanic variability. We also provide quantitative evaluations of its effectiveness in representing both distinctiveness and continuity of the profiles.

\subsection{Selecting the Color Model}
Two main considerations guided our selection of a color model for constructing the color continuum of profiles: (1) perceptual intuitiveness and (2) the ability to distinguish geometric properties across representative profiles from their colors.

To satisfy these considerations, we required a color model that aligns with human perception so that readers can intuitively associate profile features with color. Common Cartesian color spaces such as RGB and CIELAB are less suitable for this purpose, as their three components are not naturally interpretable or independently perceived by humans once combined. For example, human vision cannot easily decompose a color into separate R, G, and B components, which weakens the connection between colors and profile features. In contrast, polar color spaces such as HSV and HCL separate hue from lightness and colorfulness, which are dimensions more readily distinguished and compared by the human visual system.

The second consideration also relates to the geometric complexity of the cluster-representative profiles. In the previous section, we observed that the 80 representative profiles exhibit eight distinct shape categories. To convey this diversity, the color encoding must provide a component offering a sufficiently broad range to clearly differentiate among categories. Polar color spaces are well-suited in this respect, as their hue (H) component spans a wide range of colors from red to purple, enabling clear visual separation among distinct shapes. 

Taken together, these considerations motivate the use of polar color models. Among them, the OKLCH color model offers improved perceptual uniformity and hue consistency over earlier models, resulting in color differences that more reliably correspond to human perceived differences between profile shapes \citep{ottosson2020oklab}.

\subsection{Mapping Profile Characteristics to OKLCH Components}
\label{sec:color_component}

For each profile, we extract three numerical features: (1) the starting value $y_1$ at 200 meters depth, (2) the overall magnitude of variation, defined as $y_{max} - y_{min}$, and (3) the maximum curvature observed along the profile. These features are chosen as simple descriptors that summarize the geometric and structural characteristics of the profiles. Together with the previously defined eight shape categories, they form the basis for mapping profiles to colors by encoding the lightness (L), chroma (C) and hue (H) components in the OKLCH model. 

\subsubsection*{Lightness (L) – Encoding the Initial Level}
We encode the initial level of each profile, namely the starting value at 200 m depth $y_1$, in the lightness component. To represent profiles with higher initial temperature or salinity with greater brightness, larger $y_1$ values are mapped to higher lightness levels, whereas smaller $y_1$ values are mapped to lower lightness levels, producing darker colors. 

\subsubsection*{Chroma (C) – Encoding the Magnitude of Profile Variation}
We encode the magnitude of structural variation in the chroma component. The variation is quantified as $y_{max} - y_{min}$ over the depth range of 200 to 950 meters. To visually represent greater variation in temperature or salinity across profiles, larger variation values are mapped to higher chroma levels, producing more intense colors, whereas smaller variation values are mapped to lower chroma levels, producing softer, less saturated colors.

\subsubsection*{Hue (H) – Encoding Profile Shape and Curvature} 
We encode the shape of each cluster profile in the hue component through a two-stage design. In the first stage, each shape category is assigned to a distinct range of hue values, establishing clear visual separation at a coarse level. In the second stage, hue values are further modulated according to the curvature of individual profiles, enabling refined differentiation within categories and continuous transitions across adjacent categories.

\begin{figure}
\begin{center}
\includegraphics[width=5in]{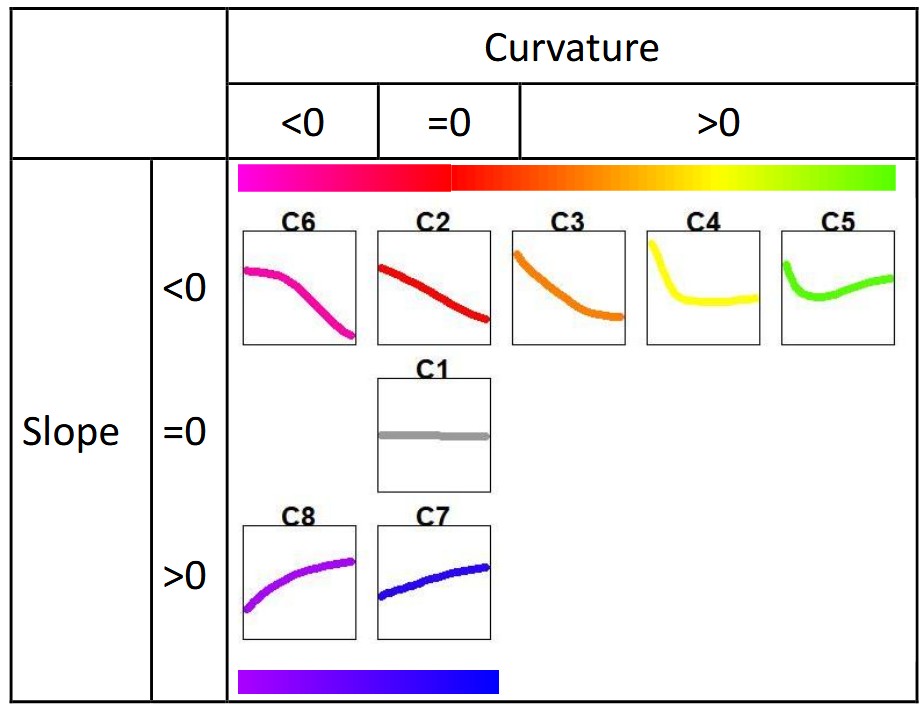}
\end{center}
\caption{Hue encoding scheme for the eight shape categories, characterized by slope and curvature.}
\label{fig:h_table}
\end{figure}

To ensure that the resulting color mapping is cognitively transparent and easy to interpret, the selection of hue ranges follows an explicit and systematic rationale. Figure~\ref{fig:h_table} illustrates the underlying logic of the first-stage design. The eight shape categories are arranged on a two-dimensional grid defined by slope (positive, zero, negative) and curvature (positive, zero, negative). The hue varies systematically with shapes and curvature, so the ordering in hue directly reflects the geometric ordering of the shape categories. The detailed rationale behind this design is explained below.

The hue configuration is anchored in the sign of the slope to reflect the climatic association between profile shapes and geographic regions. Negative-slope profiles predominantly occur in warmer regions; accordingly, categories with negative slopes are represented using warmer hues in the upper spectrum. The linear profile C2 (red, 15°–30°) serves as the central reference: moving leftward to C6 (magenta, 345°–355°), the shape becomes concave downward, whereas moving rightward it becomes concave upward. The curvature increases in magnitude from C3 (orange, 35°–85°) to C4 (yellow, 90°–105°) and ultimately to the cup-shaped profile C5 (green, 110°–120°).

In contrast, positive-slope profiles are more common in colder regions. Categories with positive slopes are therefore represented using cooler hues in the lower spectrum. The linear profile C7 (blue, 255°–285°) serves as the reference, and the hue progresses leftward along the lower spectrum toward the concave-downward category C8 (violet, 290°–310°).


The representative hue ranges established in the first stage define the conceptual structure of the hue encoding. Hue values for individual profiles are subsequently refined in the second stage.

\subsection{Numerical Local Adjustment Under Gamut Constraints}
In the OKLCH color space, not all (l, c, h) combinations lie within the displayable gamut. The assignment of these values must therefore be performed under explicit gamut constraints. Consistent with the conceptual roles defined in Section~\ref{sec:color_component}, we construct the corresponding (l, c, h) values as follows.

\subsubsection*{Step 1: Determination of Hue Values} 
Building on the representative hue ranges established in the first stage, the hue values for individual profiles are refined in the second stage. In this stage, each profile’s hue value is obtained by linearly interpolating its observed maximum curvature within the representative hue range assigned to its shape category.

\subsubsection*{Step 2: Determination of Admissible Chroma and Lightness Values}
For each hue value, we determine the admissible chroma range $[C_{\min}^h, C_{\max}^h]$ and, separately, the admissible lightness range $[L_{\min}^h, L_{\max}^h]$ that correspond to displayable colors in the OKLCH space. 

\subsubsection*{Step 3: Spline Construction of Chroma and Lightness}
We use monotone cubic Hermite splines to generate chroma values for individual profiles based on their structural variation, previously selected as the component corresponding to chroma.

Since the admissible chroma range $[C_{\min}^{h_i}, C_{\max}^{h_i}]$ varies with the hue $h_i$ of profile $i$, we construct a profile-specific spline function $s_i^C$ over the interval $[x_{min}, x_{max}]$, where $x_{min}$ and $x_{max}$ are the minimum and maximum values of $\{ x_i \}_{i=1}^{n_c}$, $n_c$ is the total number of clusters, and 
\[
x_i=max(y_i)-min(y_i)
\]
denotes the structural variation of profile $i$. The spline construction includes a set of anchor points $\{(x_{a_j}, C_{a_j})\}_{j=1}^{m}$
in addition to the boundary points $(x_{\min}, C_{\min}^{h_i})$ and $(x_{\max}, C_{\max}^{h_i})$. These anchor points prescribe chroma values at selected variation levels, allowing controlled adjustment of overall saturation in the visualization. The chroma value for profile $i$ is then obtained as  $c_i=s_i^C(x_i)$. 

For each profile, we construct an analogous monotone cubic spline $s_i^L$ to map the initial level of the profile to its lightness value. Let 
\[
z_i=y_i(200m)
\]
denote the value of profile $i$ at 200 meters depth. The lightness value is then obtained as $l_i=s_i^L(z_i)$.  

\subsubsection*{Step4: Local Gamut Adjustment for Non-Displayable Colors}
Although the chroma and lightness values for each profile are constructed within their respective admissible ranges, the resulting combinations $(h_i, c_i, l_i)$ may not necessarily lie within the displayable gamut of the OKLCH space due to the curved nature of the joint boundary. To ensure displayability, a local adjustment procedure is applied when a constructed color falls outside the gamut. Since H encodes the overall structural shape of a profile, it is kept fixed during the adjustment. Under this constraint, we search within a local chroma neighborhood of $c_i$ and select the displayable candidate that minimizes the deviation in the $(C,L)$ plane. Explicitly, candidate chroma values are explored in a local neighborhood
\[
c_i^{(k)} = c_i + \delta_k, \quad \delta_k \in \{\pm 0.1k\},
\]
where $k$ is a user-defined parameter, allowing progressively larger chroma adjustments as 
$k$ increases. Among the colors $(h_i, c_i^{(k)}, l)$ that lie within $\mathcal{G}$, 
we select the color that minimizes the deviation
\[
d_i(c,l) = |l_i - l| + |c_i - c|.
\]
That is, the final adjusted color $(h_i, c_i^*, l_i^*)$ for profile i is given by
\[
(c_i^*, l_i^*) 
= \arg\min_{\substack{c \in \{c_i^{(k)}\}, \\ (h_i, c, l) \in \mathcal{G}}}
d_i(c,l).
\]

\subsection{Profile Color Presentation}
\begin{figure}
\begin{center}

  \begin{minipage}{\textwidth}
    \centering
    
    \includegraphics[width=6.3in]{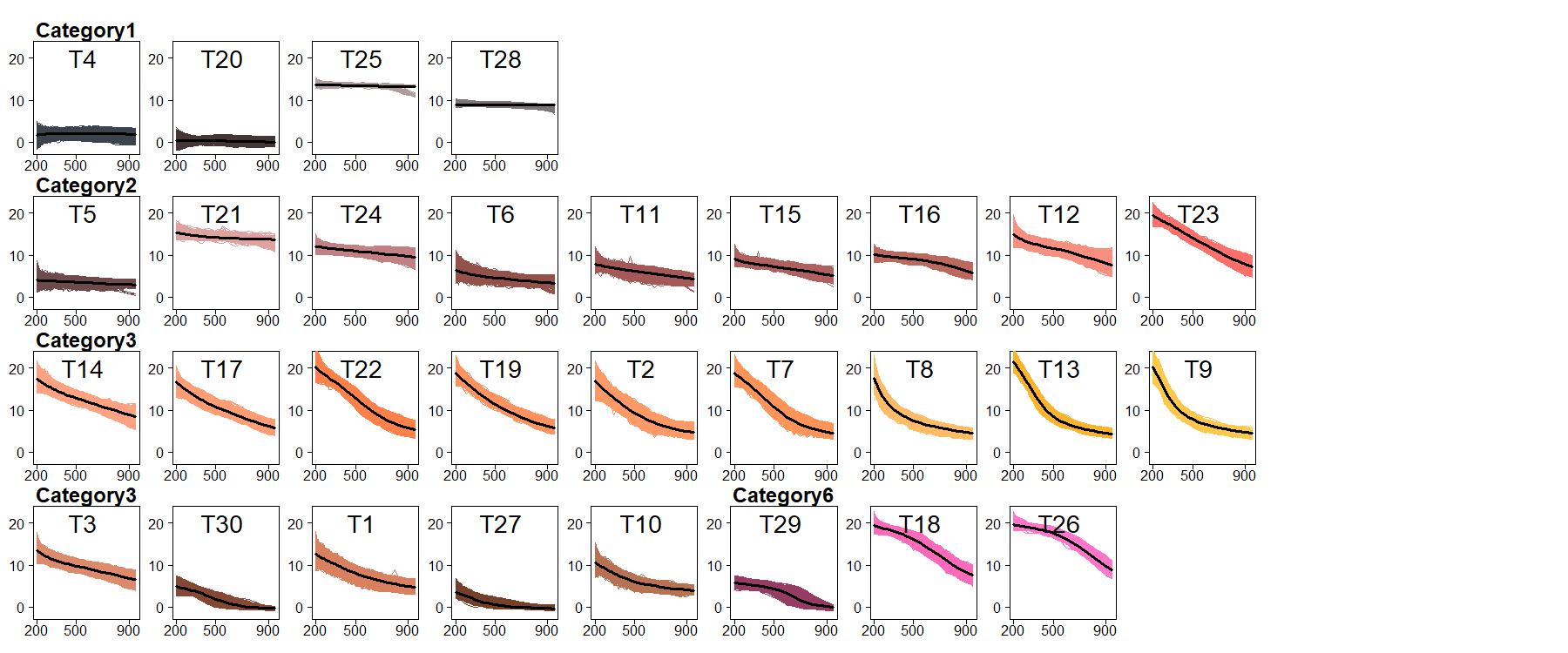}

    \vspace{-0.2cm}
    (a) Top 30 Color-Encoded Temperature Cluster Profiles
    \vspace{0.5cm}
  \end{minipage}

  \begin{minipage}{\textwidth}
    \centering
     \includegraphics[width=6.3in]{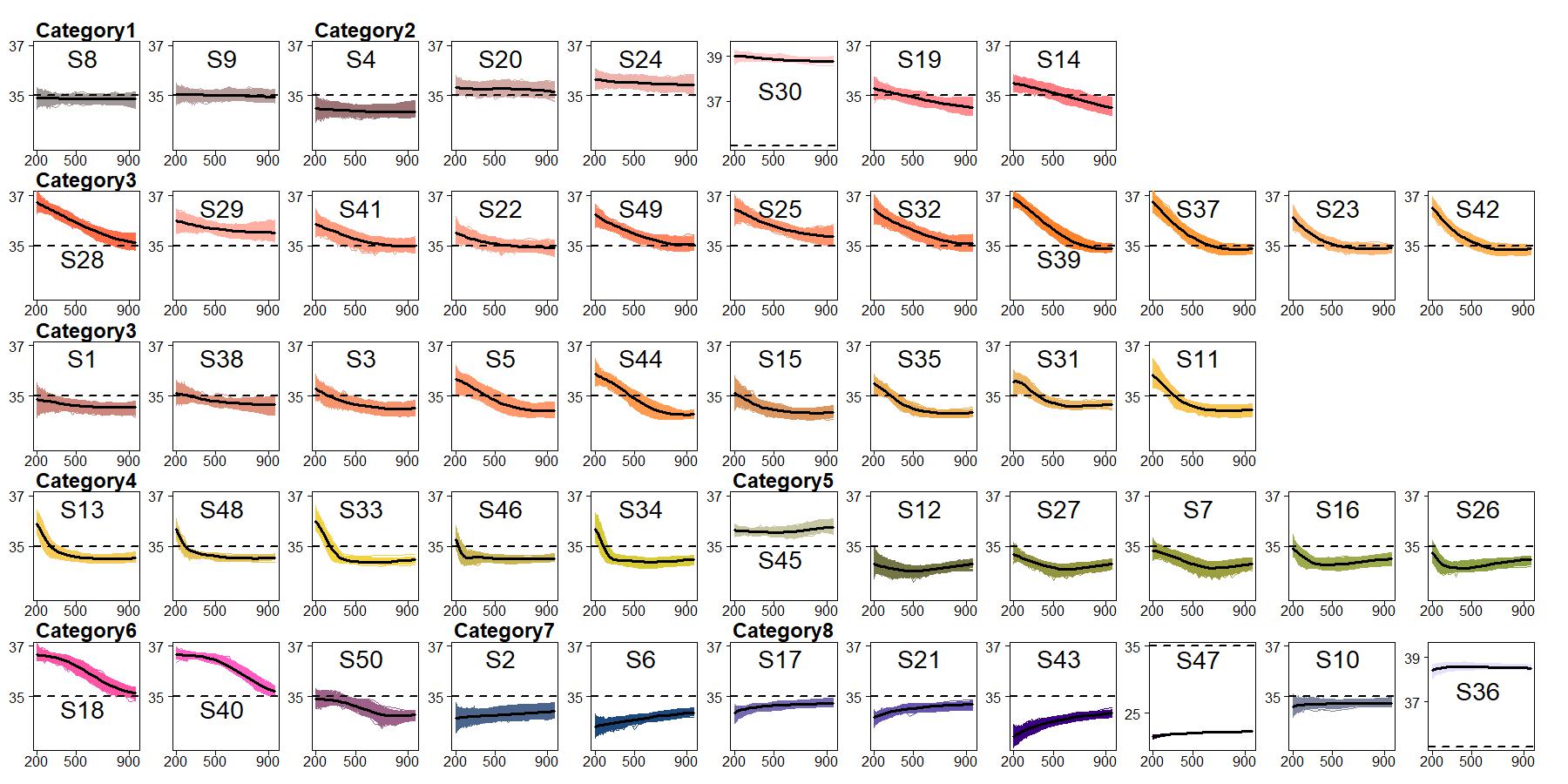}      
    \vspace{-0.2cm}
    
     (b) Top 50 Color-Encoded Salinity Cluster Profiles
   \end{minipage}
   \vspace{0.3cm}
\caption{Final color encoding of cluster-representative profiles. (a) Top 30 temperature cluster profiles. (b) Top 50 salinity cluster profiles. }
\label{fig:profiles_col}
\end{center}
\end{figure}

Figures~\ref{fig:profiles_col}(a) and~\ref{fig:profiles_col}(b) present the final color encoding of the 30 cluster-representative temperature profiles and the 50 cluster-representative salinity profiles, obtained using the proposed color construction with the following local gamut adjustment. For salinity profiles, the lightness and chroma splines are anchored at Cluster~1 by prescribing the points $(z_1,\, 0.7\,L_{\max}^{h_1})$ and $(x_1,\, 0.35\,C_{\max}^{h_1})$, respectively. For temperature profiles, the lightness anchor is specified in the same manner as for salinity, and the chroma is anchored at two points: $(x_1,\, 0.55\,C_{\max}^{h_1})$ for Cluster~1 and $(x_5,\, 0.2\,C_{\max}^{h_5})$ for Cluster~5. In both cases, the chroma adjustment parameter is set to $k=1$. These settings are chosen to achieve a balanced overall color distribution, with Cluster~1 (the largest cluster) used as the primary anchor. 

In both figures profiles are grouped by shape category. Within each category, they are ordered according to either initial profile level, magnitude of profile variation, or curvature, depending on the characteristics of that category. This arrangement is designed to highlight how differences in these three features translate to perceptible changes in the assigned colors, revealing transitions both within and across shape categories.

\subsection{Validation of the Color Construction}

We refer to the color construction developed in Sections~4.2 and~4.3 as the
SHAPE encoding, since its three components are derived directly from the
geometric shape of a cluster-representative profile rather than from a
data-driven embedding. This subsection examines whether an encoding built for
interpretability also preserves the similarity structure among profiles.

The evaluation compares pairwise distances in the profile space with the
corresponding distances in the color space. Let $D^{(P)}_{ij}$ denote the
distance between profiles $P_i$ and $P_j$, and let $D^{(C)}_{ij}$ denote the
distance between their assigned colors. If the encoding reflects the similarity
structure of the profiles, then profiles that lie close together in the profile
space should also be assigned colors that lie close together in the color space,
so that $D^{(P)}_{ij}$ and $D^{(C)}_{ij}$ vary together.

\paragraph{Color distance.}
Perceived color differences are measured in the CAM16-UCS uniform color space \citep{Li2017}.
Each assigned color is converted from its CIE XYZ coordinates to the CAM16
appearance correlates $(J, M, h)$ under a D65 white point, an adapting luminance
of $L_A = 64\,\mathrm{cd/m^2}$, a background luminance factor of $Y_b = 20$, and
average surround conditions. The correlates are then transformed to the uniform
coordinates $(J', a', b')$, and $D^{(C)}_{ij}$ is taken as the Euclidean
distance between these coordinates. This choice ensures that numerical color
differences approximate differences as perceived by a human viewer, which is the
property the encoding is intended to exploit.

\paragraph{Profile distances.}
Because no single notion of profile similarity is uniquely appropriate, we use
two complementary definitions.

The first is a \emph{basis-coefficient distance}. Each cluster-representative
profile is represented on a cubic B-spline basis with six basis functions over
the depth range 200--950~m, and $D^{(P)}_{ij}$ is the Euclidean distance between
the resulting coefficient vectors. This distance is driven by the overall
functional form of the profile and is not tied to any particular summary
feature.

The second is a \emph{feature-based distance}. Each profile is summarized by the
three descriptors introduced in Section~4.2 --- the value at 200~m, the magnitude of variation $y_{max}-y_{min}$, and the maximum curvature --- which are standardized before the Euclidean
distance is computed. These descriptors were selected in Section~4.2 on
oceanographic grounds, as the attributes of a vertical profile that carry
physical meaning and that a reader must be able to recover when interpreting the
maps. The two distances therefore represent different notions of what should be
preserved: the first asks whether the overall functional form is retained, the
second whether the physically interpretable attributes are retained. A color
encoding may succeed at one and fail at the other, and for the purpose of
reading a map it is the second that governs whether the visualization can be
used.

\paragraph{Baselines and test statistics.}
To place these results in context, we compare the SHAPE encoding with two
color mappings derived from data-driven embeddings, FPCA \citep{RamsaySilverman2005} and UMAP \citep{McInnes2018}, in which the
leading embedding coordinates are mapped onto the color components. These
methods optimize geometric objectives in the profile space and therefore
represent what is attainable when distance preservation is the explicit target.
The purpose of the comparison is not to outperform them, but to establish
whether an interpretable design can retain comparable structural consistency
while assigning explicit meaning to hue, chroma, and lightness.

Two statistics are reported. The Mantel test \citep{Mantel1967} assesses the association between
the two distance matrices, and we report the Spearman form so that the
assessment depends on the ordering of distances rather than on their scale.
Procrustes analysis \citep{Gower1975} assesses how closely the geometric configuration of the
color coordinates matches that of the profile representation, and we report the
resulting correlation $r$.

\paragraph{Scope.}
The evaluation is carried out on the 50 salinity cluster-representative
profiles, which exhibit the full range of eight shape categories; three profiles
identified as outliers are excluded, leaving 47. Because hue assignment in our
construction also reflects domain-specific rationale, geometric similarity is
not the only factor determining color differences when profiles from different
shape categories are compared. We therefore report results both for all profiles
together and separately within shape categories, the latter isolating the
geometric component of the encoding. Shape categories containing too few
profiles to support a stable comparison are omitted.

\begin{table}[t]
\caption{Agreement between profile distances and color distances. Entries are
the Mantel test Spearman correlation and the Procrustes correlation $r$; higher
values indicate closer agreement. The largest value in each row is shown in
bold. Results are based on the 47 retained salinity cluster-representative
profiles.}
\label{tab:color-validation}
\centering
\small
\begin{tabular}{llccccccc}
\toprule
 & & \multicolumn{3}{c}{Mantel (Spearman)} & & \multicolumn{3}{c}{Procrustes ($r$)} \\
\cmidrule(lr){3-5}\cmidrule(lr){7-9}
Shape category & Profile distance & FPCA & UMAP & SHAPE & & FPCA & UMAP & SHAPE \\
\midrule
\multirow{2}{*}{All}
  & Basis coefficient & 0.530 & \textbf{0.649} & 0.497 & & 0.762 & \textbf{0.785} & 0.685 \\
  & Feature           & 0.371 & 0.394 & \textbf{0.455} & & 0.682 & 0.646 & \textbf{0.721} \\
\addlinespace
\multirow{2}{*}{C2}
  & Basis coefficient & \textbf{0.612} & 0.423 & 0.551 & & \textbf{0.843} & 0.817 & 0.829 \\
  & Feature           & 0.527 & 0.602 & \textbf{0.833} & & 0.817 & 0.907 & \textbf{0.947} \\
\addlinespace
\multirow{2}{*}{C3}
  & Basis coefficient & 0.345 & \textbf{0.740} & 0.251 & & 0.692 & \textbf{0.807} & 0.641 \\
  & Feature           & 0.265 & 0.369 & \textbf{0.437} & & 0.679 & 0.681 & \textbf{0.703} \\
\addlinespace
\multirow{2}{*}{C4}
  & Basis coefficient & 0.467 & 0.309 & \textbf{0.673} & & 0.841 & 0.744 & \textbf{0.938} \\
  & Feature           & 0.685 & 0.552 & \textbf{0.842} & & 0.836 & 0.746 & \textbf{0.980} \\
\addlinespace
\multirow{2}{*}{C5}
  & Basis coefficient & 0.261 & 0.430 & \textbf{0.491} & & 0.761 & 0.777 & \textbf{0.831} \\
  & Feature           & 0.321 & 0.127 & \textbf{0.382} & & 0.717 & 0.533 & \textbf{0.792} \\
\addlinespace
\multirow{2}{*}{C7}
  & Basis coefficient & 0.886 & 0.886 & \textbf{0.943} & & 0.951 & 0.918 & \textbf{0.984} \\
  & Feature           & 0.771 & 0.771 & \textbf{1.000} & & 0.950 & 0.897 & \textbf{0.993} \\
\bottomrule
\end{tabular}
\end{table}

\paragraph{Results for all profiles.}
Table~\ref{tab:color-validation} reports the results. When all profiles are
considered together and similarity is measured by the basis-coefficient
distance, the embedding-based mappings show the stronger agreement, with Mantel
correlations of 0.530 for FPCA and 0.649 for UMAP against 0.497 for SHAPE; the
Procrustes correlations follow the same ordering. This is the expected outcome,
since only the embeddings are constructed to preserve distances. The magnitude
of the gap is nevertheless modest, indicating that the SHAPE encoding retains a
substantial part of the similarity structure even though nothing in its design
targets that structure.

The ordering reverses under the feature-based distance, and the manner of the
reversal is informative. The agreement achieved by the embedding-based mappings
falls sharply when the target changes, from 0.530 to 0.371 for FPCA and from
0.649 to 0.394 for UMAP, whereas the SHAPE encoding is largely unaffected,
moving only from 0.497 to 0.455. The Procrustes correlations behave the same
way. What the embeddings preserve is therefore the overall functional geometry
of the profiles, and that geometry is not equivalent to the physically
interpretable attributes: a mapping can rank highly on the first criterion while
losing much of the second. The consequence is practical rather than technical.
If the colors on a map do not preserve differences in initial level, magnitude
of variation, and curvature, a reader cannot recover those quantities from the
map, whatever the mapping's overall distance-preservation properties may be.
This is the case for constructing the encoding directly from the profile
features rather than from an embedding.

\paragraph{Results within shape categories.}
Restricting the comparison to profiles within the same shape category removes
the influence of the categorical hue assignment and isolates the finer,
continuous part of the encoding. The SHAPE model performs markedly better under
this restriction. It attains the highest agreement in every category under the
feature-based distance and in categories C4, C5, and C7 under the
basis-coefficient distance, for both the Mantel and Procrustes statistics. The
gains are largest in the categories with pronounced curvature: in C4, the Mantel
correlation reaches 0.673 under the basis-coefficient distance against 0.467 for
FPCA and 0.309 for UMAP, and the Procrustes correlation reaches 0.938 against
0.841 and 0.744. In C2 the corresponding feature-based Mantel correlation is
0.833, well above 0.527 for FPCA and 0.602 for UMAP, with the same pattern in
the Procrustes results.

The exception is C3, where UMAP retains a clear advantage under the
basis-coefficient distance (0.740 against 0.251). C3 is the largest and most
heterogeneous category, spanning a wide range of initial levels and variation
magnitudes within a single convex-decreasing form, so within-category color
differences are constrained by the hue interval allocated to the category. This
constraint is the cost of preserving a globally interpretable hue ordering.

Two limitations should be noted when reading the within-category results. The
within-category permutation tests use a smaller number of permutations than the
all-profile test, so the associated $p$-values are correspondingly coarse.
Furthermore, categories represented by only a few profiles yield correlations
based on a small number of pairwise distances, which are unstable and in some
cases attain their maximum for every method; such categories are excluded from
the table.

\paragraph{Summary.}
Taken together, the results separate two criteria that are easily conflated.
Judged on the preservation of overall functional geometry with all profiles
pooled, the embedding-based mappings hold a modest advantage, as their
objectives lead one to expect, and the SHAPE construction remains close behind.
Judged on the preservation of the physically interpretable profile attributes,
the embeddings lose a substantial part of the structure while the SHAPE
construction retains it, and this holds both across all profiles and, more
strongly, within shape categories, where the SHAPE construction matches or
exceeds the embeddings in most comparisons. Since it is the second criterion
that determines whether a reader can recover profile characteristics from the
colors on a map, these results support constructing the encoding directly from
the profile features. The resulting design gives up little in overall structural
consistency and in return assigns explicit and stable meaning to lightness,
chroma, and hue.

\section{Global Patterns Visualization}
\label{sec:global_vis}

\subsection{Temperature and Salinity Global Patterns}

\begin{figure}
\begin{center}
\includegraphics[width=5in]{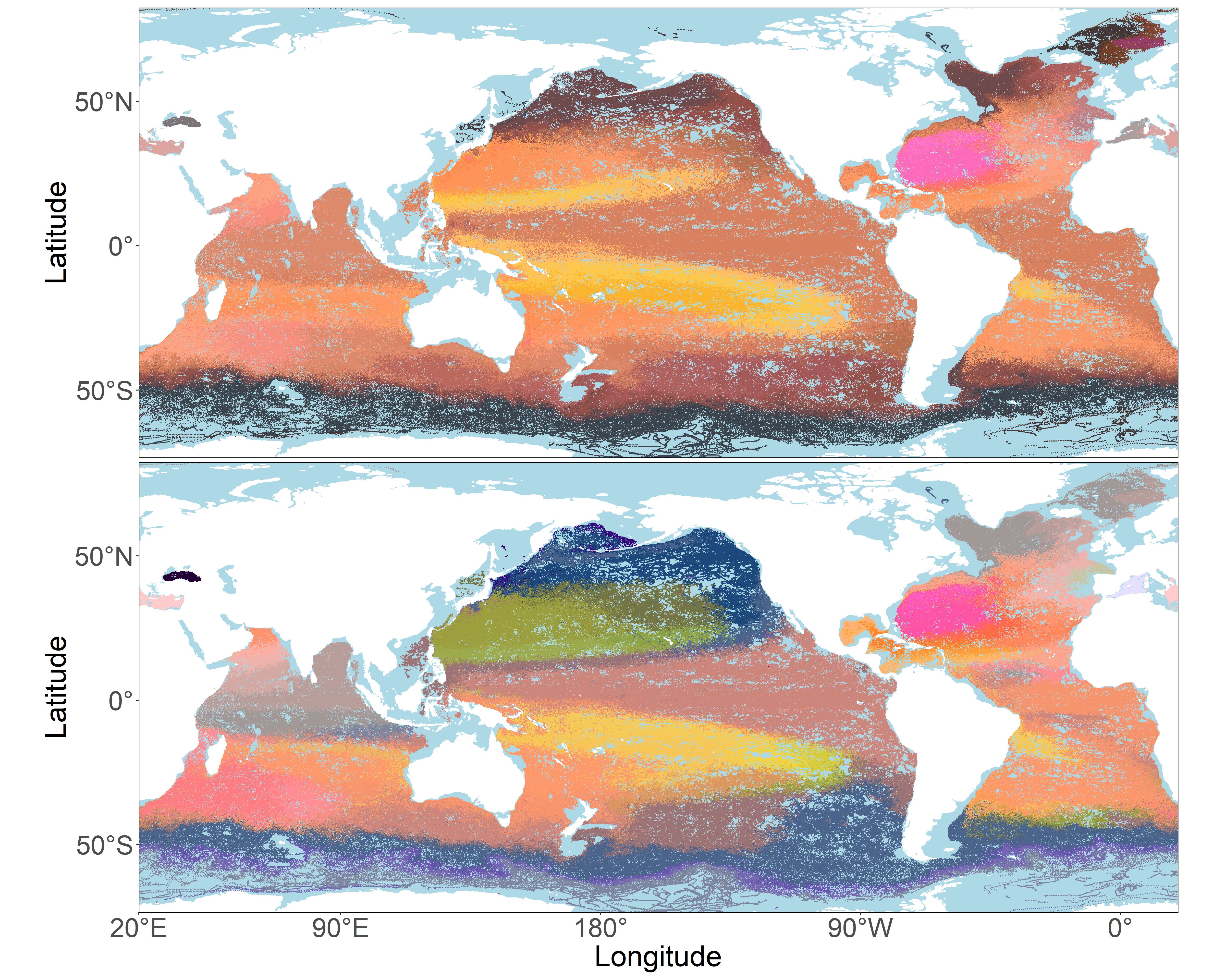}
\caption{Global distributions of (a) vertical temperature patterns and (b) vertical salinity patterns.
In each grid cell, the color represents the profile shape of the predominant cluster, determined by majority vote and encoded using the color scheme introduced in Section~4.}
\label{fig:worldmap_cluster}
\end{center}
\end{figure}

Following the color encoding introduced in the previous section, Figure~\ref{fig:worldmap_cluster}(a) and~\ref{fig:worldmap_cluster}(b) present the global vertical structures of temperature and salinity over the ten-year period from 2014 to 2023. Each map displays approximately 800,000 colored oceanic grid cells at a 0.1° × 0.1° resolution, with the color of each cell determined by the predominant cluster observed at that location during the decade.

%
%
%
%
%
%
%
%

To guide the interpretation of the maps, this section presents examples from the three major ocean basins, focusing on the vertical structures that can be inferred from the colors in the corresponding regions. For clarity and brevity, only Figure~\ref{fig:worldmap_cluster}(b), the salinity map, is used for illustration.

\subsubsection*{The Atlantic Ocean}
In the Atlantic Ocean, most regions are shaded orange (C3), indicating a convex profile: the salinity value decreases more rapidly at shallow depths and more slowly at greater depths. However, the lightness, saturation, and hue of the orange tones vary across regions. The North Atlantic displays a brighter and more saturated orange compared to the South Atlantic, suggesting higher salinity values followed by a greater overall decrease. Moreover, the orange hue in the North tends toward red (C2), whereas in the South it leans more toward yellow (C4). This color difference suggests a smaller curvature in the North, meaning that the salinity decreases more steadily with depth in the North, resulting in a profile that is closer to linear, in contrast to the more curved decline observed in the South. 

There is also a distinct magenta (C6) area in the western North Atlantic, where the color vividness indicates high salinity at $200$m depth and a pronounced decline. The magenta tone corresponds to a concave profile: the salinity value decreases gradually near the surface, followed by a sharper decrease at greater depths.

\subsubsection*{The Pacific Ocean}

In the Pacific Ocean, the wide range of colors reflects a greater diversity in profile shapes. The equatorial region is presented by a dull, pale red with an orange tint. This color suggests moderate salinity values that decrease only slightly with depth, following a convex but nearly linear pattern. In the subtropical regions, green (C5) in the North indicates a salinity profile that decreases and then slightly increases at intermediate depths, whereas yellow (C4) in the South shows a decrease followed by a flat profile without an upward trend. Within the same color category, subtle differences in color reflect finer variations in profile shape. For example, within the northern green zone, yellow-green areas reflect smoother profiles with smaller curvature, whereas purer green areas indicate more abrupt decreases followed by increases, with larger curvature.

Both polar regions are presented with blue (C7) and purple (C8) tones, indicating that salinity values increase with depth. Salinity at 200 meters depth is low in both regions, but the North shows darker and more saturated colors, suggesting even lower salinity at 200 meters depth followed by a stronger increase. A broad three-layer purple band encircling Antarctica reflects concave upward profiles rather than linear ones. This layered structure within the purple color family further highlights the smooth transitions among the three corresponding cluster profile structures. A small purple region near the Bering Sea in the Arctic shows a similar concave structure, but with a deeper and more saturated purple tone.

\subsubsection*{The Indian Ocean}
In the Indian Ocean, the western region from the Gulf of Oman to the equator shows a transition from bright orange to pale pink, indicating high salinity and a strong decline with depth in the north, which becomes less pronounced toward the equator. In the eastern region, from the Bay of Bengal to just south of the equator, colors shift from bright gray to light purple. This transition begins near the equator, where the colors show a subtle darkening, indicating a slightly lower salinity toward the South. The gray area (C1) corresponds to nearly flat profiles, with salinity values showing little variation with depth. Similar to the gray region, the light purple area also exhibits a nearly flat profile, but with a slight increase in salinity early in the profile.

In the southwestern basin, bright red indicates high salinity at 200 meters depth  followed by a strong, near-linear decline. Other regions appear in pink, orange, and yellow tones, corresponding to previously described profile shapes.

\subsection{Temperature-Salinity Associations}

Each of Figure~\ref{fig:worldmap_cluster}(a) and~\ref{fig:worldmap_cluster}(b) displays the global distribution of a single oceanic variable, temperature and salinity, respectively. Comparing the two maps yields an additional observation: the spatial contours of their respective cluster boundaries are remarkably similar, suggesting a potential connection between temperature and salinity patterns. Motivated by this observation, we explore the interplay between these two variables. Specifically, we want to understand how their joint patterns vary across different oceanic regions, such as the typical salinity structures associated with different types of temperature profiles.

\begin{figure}
    \centering

    \begin{minipage}[b]{0.49\textwidth}
        \centering
        \includegraphics[width=\linewidth]{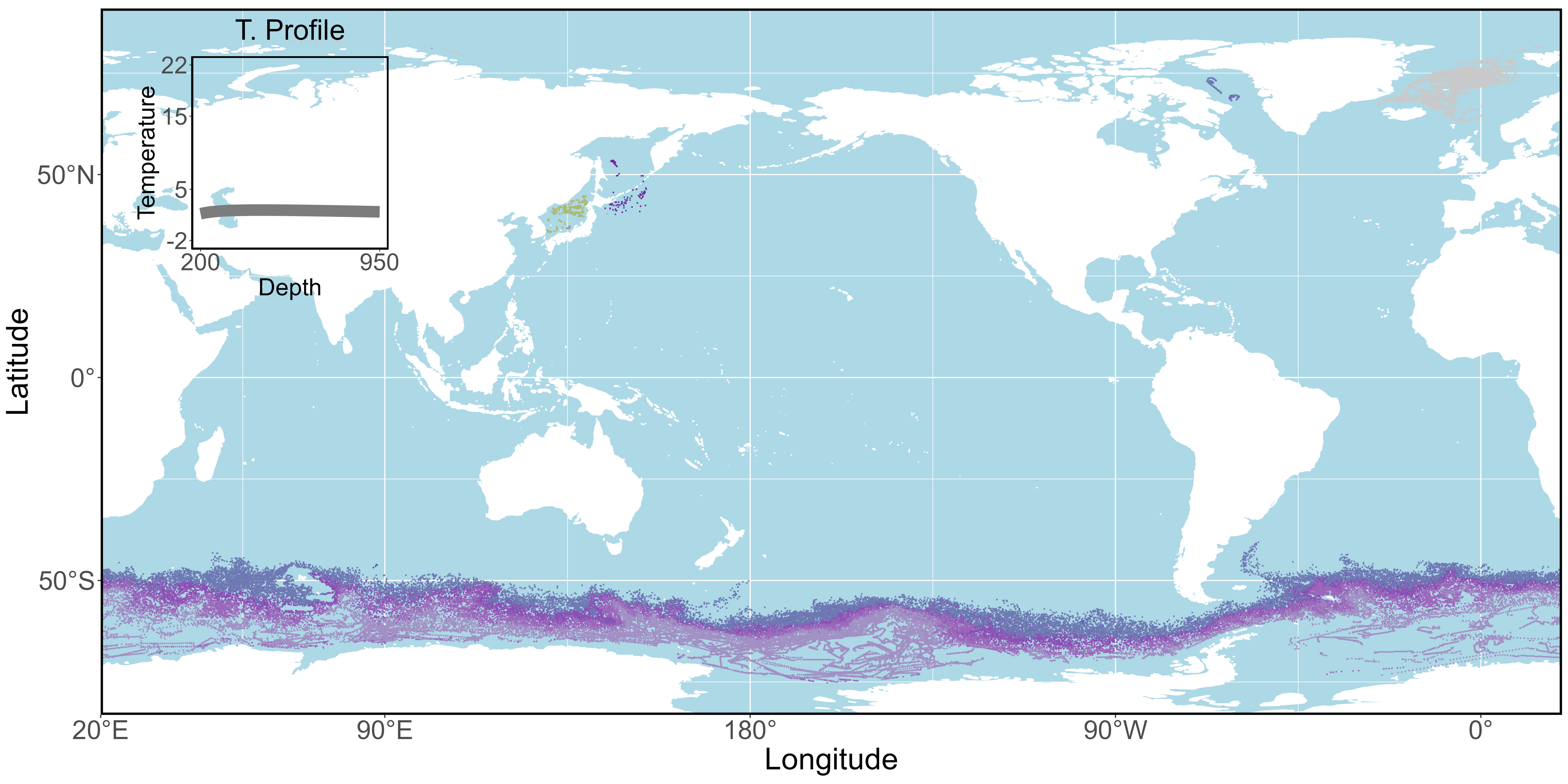}
        
        \vspace{-0.3cm}
        (a)  Flat

    \end{minipage}
    \hfill
    \begin{minipage}[b]{0.49\textwidth}
        \centering
        \includegraphics[width=\linewidth]{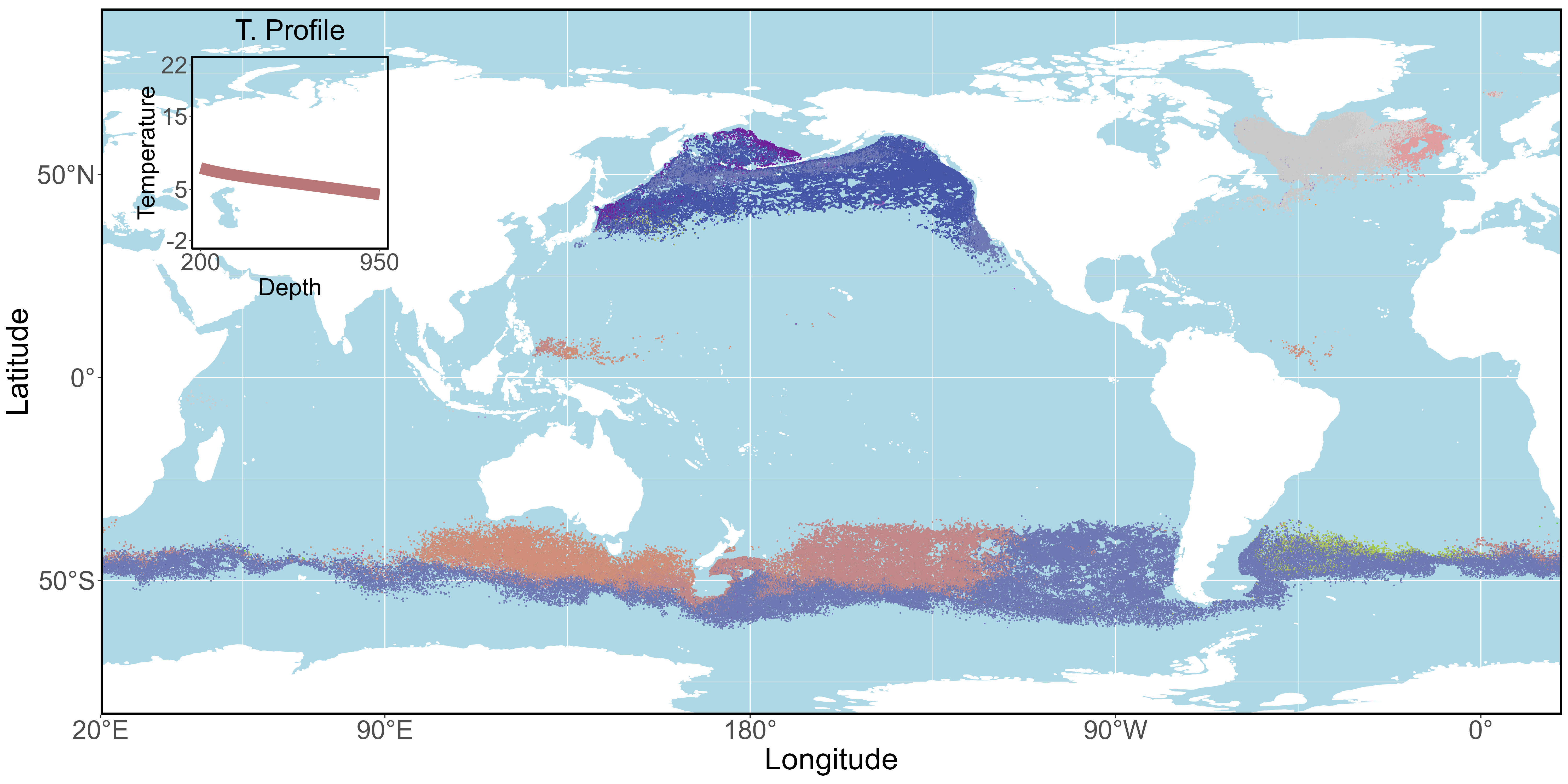}
        
        \vspace{-0.3cm}
        (b)  Negative Slope, Cold
    \end{minipage}

    \vspace{0.8cm}

    \begin{minipage}[b]{0.49\textwidth}
        \centering
        \includegraphics[width=\linewidth]{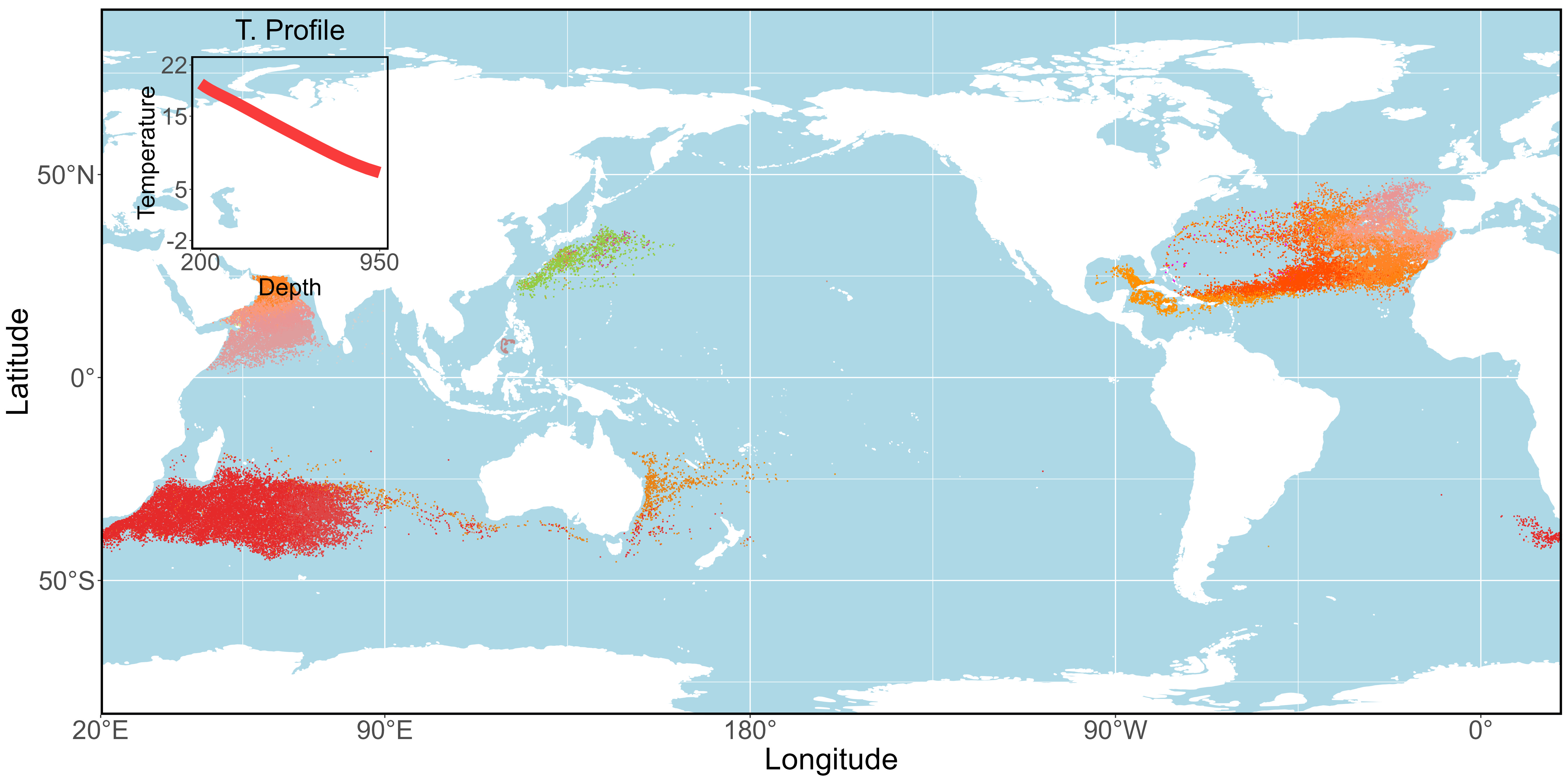}
        
        \vspace{-0.3cm}
        (c)  Negative Slope, Warm
    \end{minipage}
    \hfill
    \begin{minipage}[b]{0.49\textwidth}
        \centering
        \includegraphics[width=\linewidth]{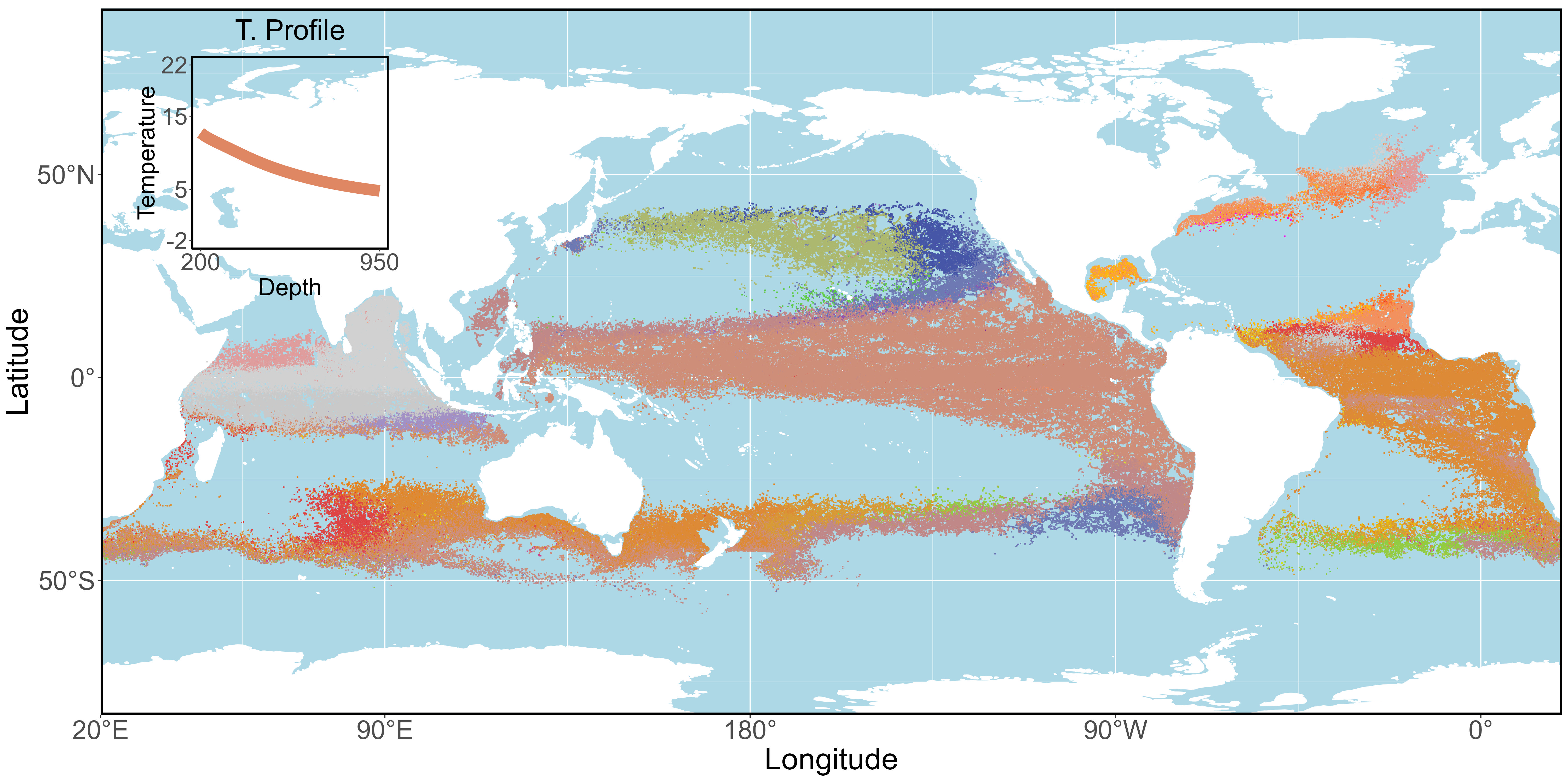}
        
       \vspace{-0.3cm}
        (d)  Convex, Cold
    \end{minipage}

    \vspace{0.8cm}

    \begin{minipage}[b]{0.49\textwidth}
        \centering
        \includegraphics[width=\linewidth]{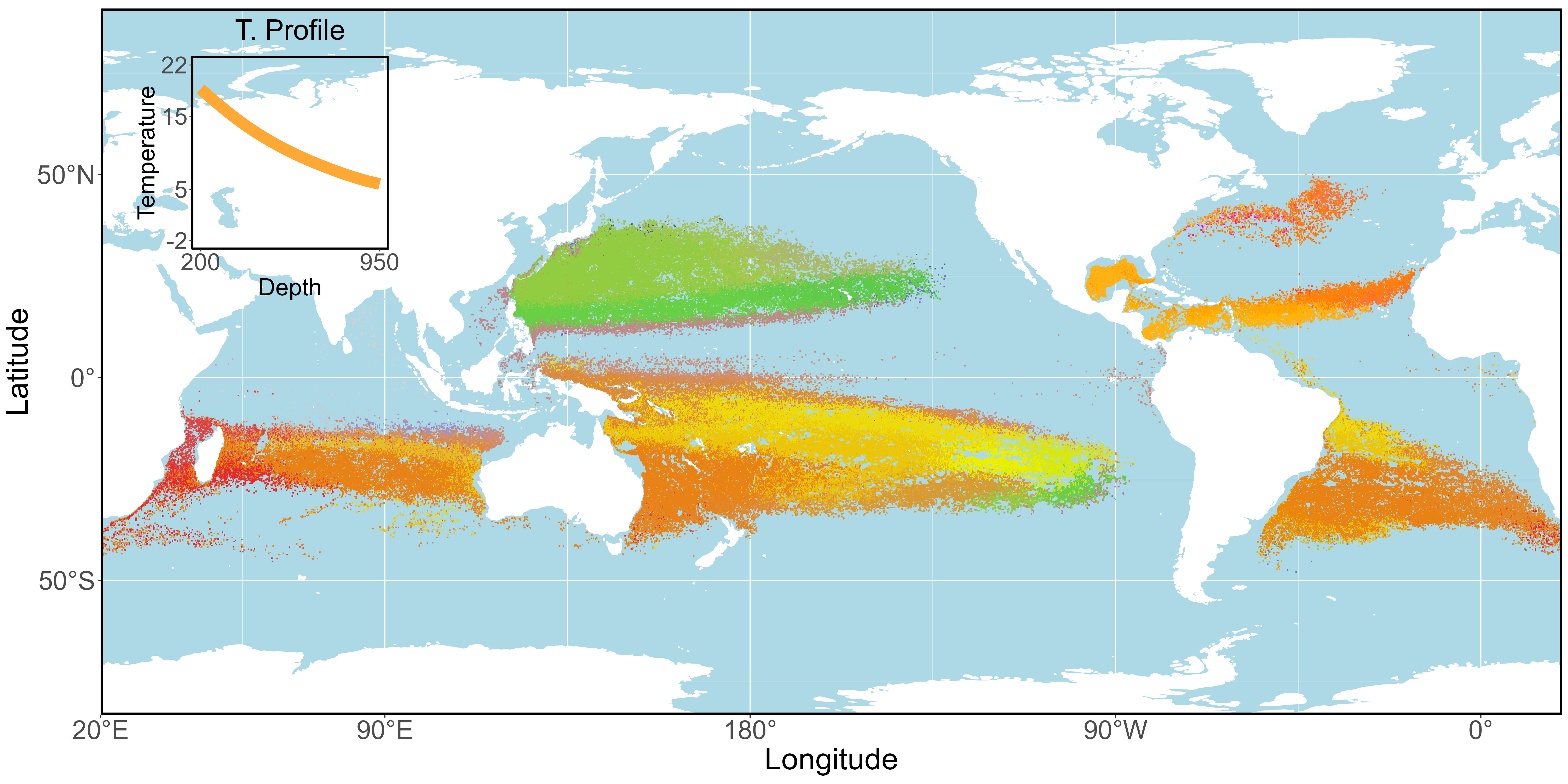}

       \vspace{-0.3cm}
        (e)  Convex, Warm
    \end{minipage}
    \hfill
    \begin{minipage}[b]{0.49\textwidth}
        \centering
        \includegraphics[width=\linewidth]{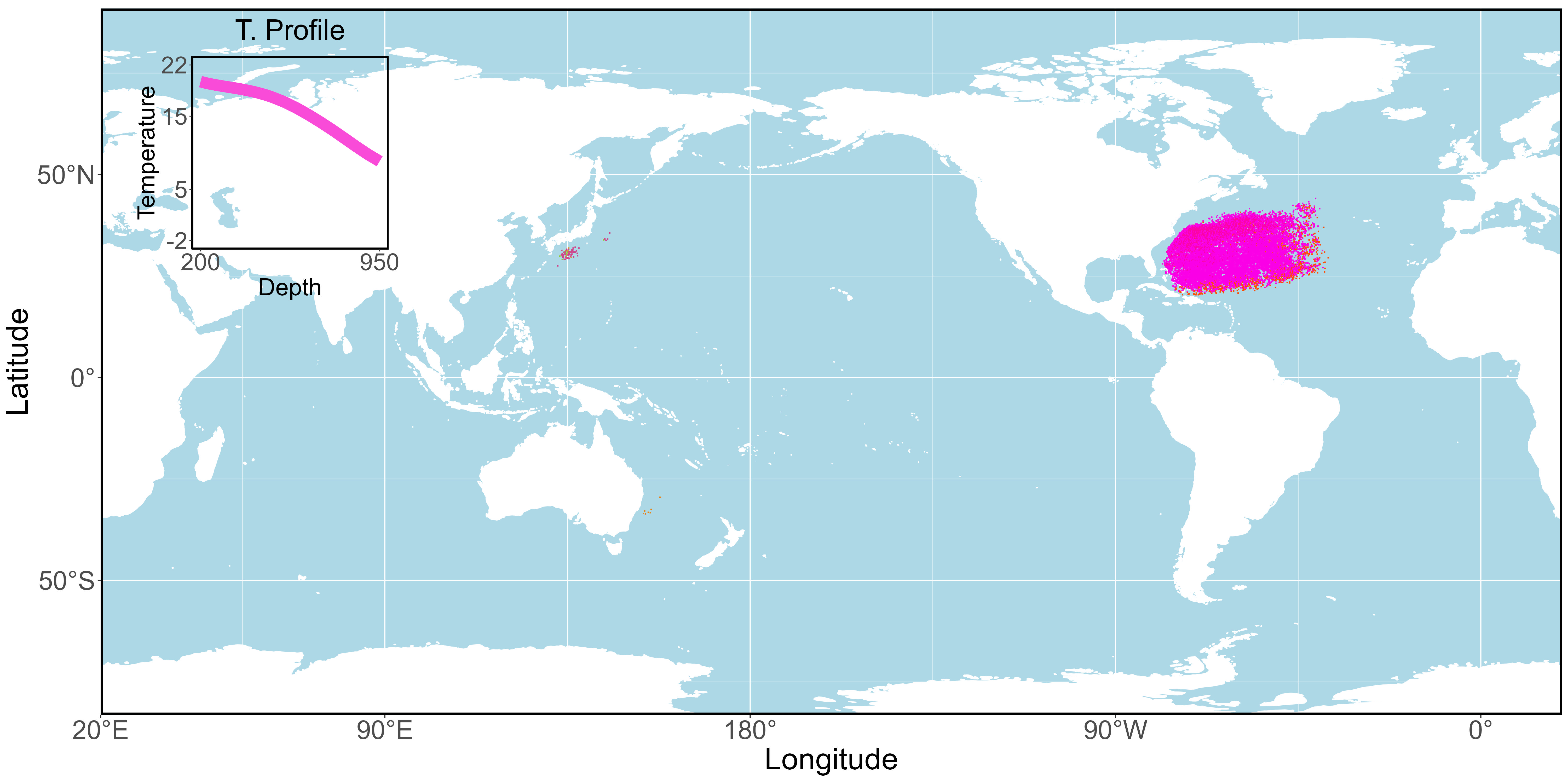}

       \vspace{-0.3cm}
        (f)  Concave
    \end{minipage}

    \caption{Distributions of salinity vertical patterns given representative temperature profiles. Insets show the temperature profile type for each panel: (a) Flat, (b) Negative slope with low temperature, (c) Negative slope with warm temperature, (d) Convex with low temperature, (e) Convex with warm temperature, (f) Concave. Grid cell colors indicate the predominant salinity cluster observed in that region.}
    \label{fig:ts_assoc}
\end{figure}

Figure~\ref{fig:ts_assoc} presents a series of maps that illustrate the co-occurring structure of temperature and salinity across different parts of the ocean. Each panel corresponds to a representative type of temperature profile (shown in the inset) and highlights the oceanic regions where this temperature pattern is observed. Within these regions, each grid cell is colored according to the predominant salinity cluster identified at that location. This figure shows how salinity patterns vary in relation to different temperature profile types.

Panel (a) presents flat and low temperature profiles, primarily occurring in the Southern Ocean near Antarctica. The corresponding salinity values are also low, increasing with depth either linearly (C7, blue) or convexly (C8, purple). Panel (f) presents concave temperature profiles with high initial values and strong vertical gradients, observed in the northwestern Atlantic. The magenta (C6) coloring in this region, which shares a similar tone with the temperature profiles, indicates that salinity also follows a pronounced concave pattern.

For panels (b) through (e), similar patterns of correspondence between temperature and salinity are observed. For instance, the representative temperature profile presented in panel (e) is more pronounced than that in panel (d), with both higher initial temperatures and strong vertical gradients. Correspondingly, the salinity patterns in panel (e) are represented by brighter and more saturated colors than those in panel (d), suggesting more pronounced salinity structure.

Together, these observations point to a strong structural correspondence between temperature and salinity in the global ocean. Regions with low temperatures generally also exhibit low salinity. Likewise, regions in which temperature shows strong vertical gradients, salinity tends to vary strongly with depth as well.

\newpage
\section{Conclusion and Discussion}
We present a framework for studying high-dimensional spatial patterns in large-scale and complex systems, developed and illustrated through the analysis of vertical temperature and salinity structures in the global ocean using Argo data. This framework consists of three components. First, we retain the full depth-dependent structure of individual profiles, in contrast to traditional approaches that rely on physical thresholds or regions defined by expert knowledge. This data-driven approach preserves the detailed information inherent in each profile. Second, we apply the randomized self-updating process (rSUP) clustering algorithm to identify representative characteristics that summarize how ocean profiles vary across different regions. We emphasize clustering as a form of dimension reduction and a discrete approximation to the global patterns. Finally, we present a color encoding design that integrates information from individual clusters to reveal both spatial patterns and gradual transitions between them. This encoding emphasizes interpretability and makes it possible to move from a discrete approximation to a visually coherent representation. By bringing together these three components, the framework enables a comprehensive and interpretable visualization of large-scale vertical structures across the global ocean.

\begin{figure}
\begin{center}

  \begin{minipage}{\textwidth}
    \centering
    \includegraphics[width=6.4in]{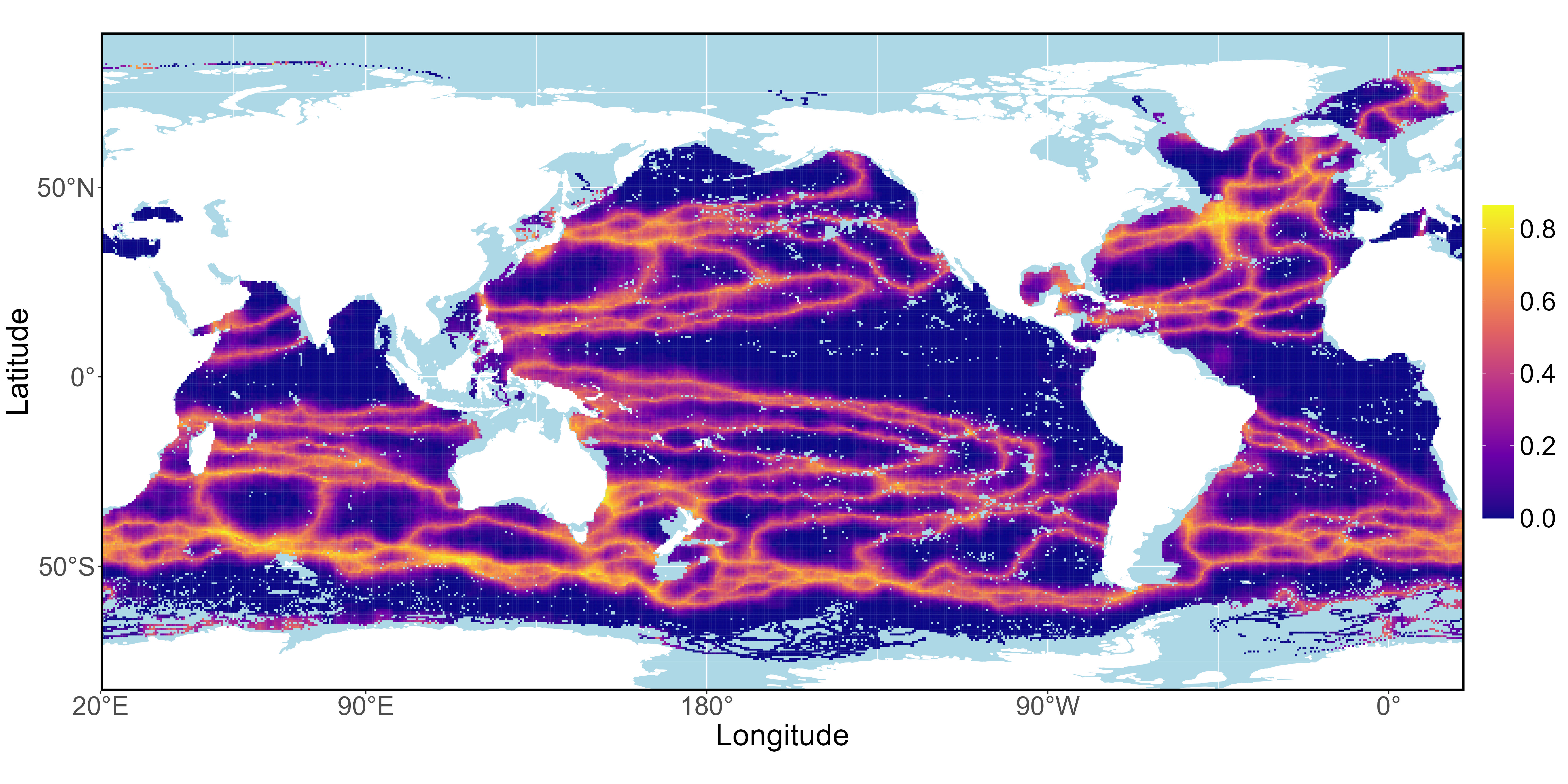}

    \vspace{-0.1cm}
 
    (a) Temperature Predominant Cluster Variability
   \end{minipage}

   \vspace{0.8cm}
   
   \begin{minipage}{\textwidth}
   \centering
   \includegraphics[width=6.4in]{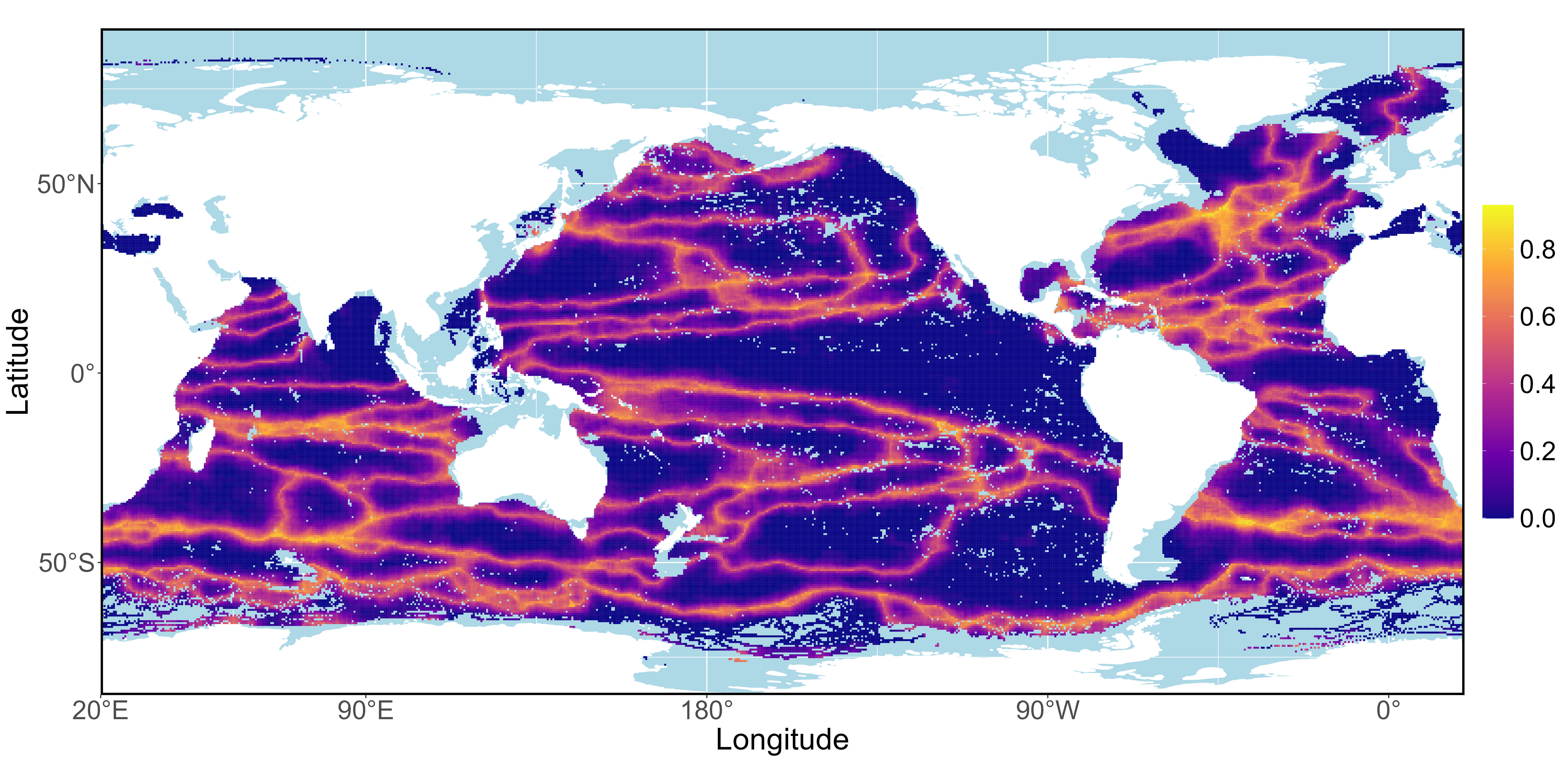}
   
   \vspace{-0.1cm}
   (b)  Salinity Predominant Cluster Variability
   \end{minipage}

   \vspace{0.5cm}
   
\caption{Spatial distribution of predominant cluster variability for (a) temperature and (b) salinity over a ten-year period. Higher values (in yellow) indicate greater temporal variability.}
\label{fig:worldmap_majC_change}
\end{center}
\end{figure}

While our primary focus was on the spatial distribution of vertical temperature and salinity patterns over a ten-year period, we also conducted preliminary explorations of how these patterns vary over time. Figure~\ref{fig:worldmap_majC_change}(a) and~\ref{fig:worldmap_majC_change}(b) show the predominant cluster variability for each grid cell over the ten-year period. Larger 5° × 5° grids were used in order to obtain sufficient monthly data within each spatial unit. For each grid, the predominant cluster variability is computed as the proportion of months (over a 10-year period, totaling 120 months) in which the monthly predominant cluster differs from the overall ten-year predominant cluster. This proportion reflects the temporal variability of the dominant vertical patterns in a region. A value close to 0 indicates strong consistency across time, while a value close to 1 often indicates that the long-term predominant cluster is rarely dominant in individual months, highlighting substantial temporal variability in the vertical profile structure of that region.

Figure~\ref{fig:worldmap_majC_change}(a) and~\ref{fig:worldmap_majC_change}(b) show that some regions exhibit strong temporal consistency, such as the equatorial Pacific, and some regions display high variability, such as the subpolar North Atlantic. In addition, a series of filament-like contours can be seen throughout the maps, marking areas of high variability. These contours correspond closely to the boundaries between previously identified clusters. The instability observed along these boundaries may arise either from gradual transitions between adjacent profile patterns, reflecting spatial proximity, or from systematic temporal processes linked to seasonal, interannual, or broader climate-related influences. Although the mesopelagic zone examined in this study is generally less affected by seasonal and climate-related influences than the surface ocean, our preliminary analysis showed subtle links. These observations, however, may reflect limited temporal sampling or other confounding factors, and thus should be interpreted with caution. While the results are not conclusive, the proposed clustering framework may be applied to a systematic investigation of temporal dynamics. This could help explore long-term changes in ocean structure in response to climate variability.

\section*{Acknowledgments}
The authors acknowledge the use of OpenAI's ChatGPT to improve the English
presentation of the manuscript, and of Anthropic's Claude to draft portions of
Section 4.5. All AI-assisted text was reviewed, verified against the underlying
analysis, and edited by the authors, who take full responsibility for the
content of the manuscript.

\section*{Data Availability Statement}
The data that support the findings of this study are available from the Argo Data Management website (\url{http://www.argodatamgt.org}). These data were collected by the international Argo program (\url{https://argo.ucsd.edu}), which is supported by national programs and made freely available.



\bibliographystyle{plainnat}
\bibliography{argo}

\end{document}